\documentstyle[prl,aps,preprint,tighten,floats,aps,epsf,psfig,axodraw]{revtex}
\def\simlt{\stackrel{<}{{}_\sim}}

\def\be{\begin{equation}}
\def\ee{\end{equation}}
\def\bear{\begin{eqnarray}}
\def\eear{\end{eqnarray}}
\def\beqn{\begin{eqnarray}}
\def\eeqn{\end{eqnarray}}

\def\beq{\begin{equation} }
\def\eeq{\end{equation} }

\def\ben{\begin{eqnarray} }
\def\een{\end{eqnarray} }

\def\mod#1{{\rm (mod~2)} }

\def\Tr{{\rm Tr}}

\def\simlt{\stackrel{<}{{}_\sim} }

\def\Tr{{\rm Tr}\,}

\def\ln{{\rm ln}\,}

\begin{document}
\draft
\preprint{\vbox{\baselineskip=12pt
\rightline{CERN-TH/98-243}
\vskip0.2truecm
\rightline{UPR-0811-T}
\rightline{IEM-FT-177/98}
\vskip0.2truecm
\rightline{UM-TH-98/17}
\vskip0.2truecm
\rightline{hep-ph/9807479}}}

\title{Physics Implications of Flat Directions in Free Fermionic Superstring
Models I: Mass Spectrum and Couplings} 
\author{G. Cleaver${}^{\dagger}$\footnote{Present address: Center for Theoretical Physics, Texas A \& M University, College Station, Texas 77843-4242, USA}, M. Cveti\v c${}^{\dagger}$, 
J. R. Espinosa${}^*$, L. Everett${}^{\dagger}$\footnote{Present address: Randall Laboratory of Physics, University of Michigan, Ann Arbor, MI 48109, USA }, 
P. Langacker${}^{\dagger}$, and J. Wang${}^{\dagger}$}
\address{${}^{\dagger}$Department of Physics and Astronomy \\ 
          University of Pennsylvania, Philadelphia PA 19104-6396, USA \\
${}^*$CERN, TH Division\\
CH-1211 Geneva 23, Switzerland\\
}
\maketitle
\begin{abstract}
From the ``top-down'' approach we investigate physics implications of 
the class of $D$- and $F$- flat directions
formed from non-Abelian singlets
which are proven flat
to all orders in the nonrenormalizable superpotential, 
 for  a  prototype  quasi-realistic free fermionic
string model  with the standard model gauge group and three families
(CHL5). 
These flat directions have at least
an additional $U(1)'$ unbroken at the string scale.  
For each flat
direction, the complete set of 
effective mass terms and effective trilinear superpotential
terms  in the observable sector  are computed  to all orders in the
VEV's of the fields in the 
flat direction. The ``string selection-rules''  disallow a large number
of couplings  allowed by gauge invariance, 
resulting in a 
massless spectrum  with a  large number
of  exotics, in most cases
excluded by experiment, thus signifying a generic
flaw of these models. Nevertheless, the resulting trilinear couplings of the
massless spectrum possess a number of interesting features
 which
we analyse for  two representative flat directions: for
the fermion texture;  baryon- and lepton-number violating
couplings; $R$-parity breaking; non-canonical $\mu$ terms;
 and the possibility of electroweak and intermediate scale  symmetry breaking
scenarios
for 
$U(1)'$.
The gauge coupling predictions are obtained in the electroweak
scale case.
Fermion masses possess 
$t-b$ and $\tau-\mu$
 universality, with the string scale  Yukawa couplings $g$
and $g/\sqrt{2}$,
respectively.
Fermion textures
are present  for certain flat directions, but only
in the down-quark sector.
Baryon- and lepton- number violating couplings can trigger
proton-decay,
 $N-{\bar N}$ oscillations,  leptoquark interactions 
and  $R$-parity violation, leading to the absence of a stable LSP.
\end{abstract}
\vskip2cm
\newpage
\section{Introduction}

At present, there are several challenges to be faced in the investigation
of the implications of superstring theory for physics beyond the Standard
Model (SM).  A primary obstacle is the degeneracy of string vacua; a large
number of string models have been constructed, with as yet no fully
realistic model.  There is also
no satisfactory scenario for supersymmetry breaking in string theory
either at the level of the world-sheet dynamics or at the level of the
effective theory, and hence no way to break supersymmetry in string models
without introducing new parameters.  Both issues are hoped to have a
resolution with greater understanding of nonperturbative string dynamics.  

Our strategy is to take a more modest view by restricting our
consideration to a class of string models of
 perturbative heterotic string vacua~\cite{CY,lat,ABKW,KLST,DHVW} which
have the ingredients of the minimal supersymmetric standard model (MSSM),
and thus the potential to be realistic.  Such quasi-realistic models have been
constructed in a weakly coupled heterotic superstring theory in a variety of
constructions~\cite{orbifolds,calabiyau,NAHE,FNY1,AF1,chl}.  We consider a
 class of free fermionic models~\cite{NAHE,FNY1,AF1,chl}
 which have $N=1$ supersymmetry, the SM gauge group as a part of the gauge
structure, and candidate fields for the three generations of quarks and
leptons as well as two electroweak Higgs doublets.  These models also
possess gauge coupling unification at $M_{\rm String}\sim 5 \times
10^{17}$ GeV~\cite{Kap} without a gauge group
unification; this scale differs by an order of magnitude from the
unification scale obtained by extrapolating from the observed low-energy
values of the gauge couplings assuming the minimal particle content of the
MSSM. (For a review of the properties of string models, see
Ref.~\cite{dienes} and references therein.)

This class of models share a number of generic features.  Their gauge
 structures contain at tree level an additional non-Abelian ``hidden" sector
gauge group as well as a number of Abelian gauge groups, one of them
generically anomalous.  The SM hypercharge is determined as a linear
combination of the non-anomalous $U(1)$'s of the model (or perhaps of the
$U(1)$'s that arise when the hidden sector gauge group is broken).  In
addition to the MSSM fields, the particle content typically includes a
number of fields which are nontrivial representations under the SM
(observable sector) gauge group or the non-Abelian hidden sector gauge
group (or both), as well as a number of non-Abelian singlet
fields.  Most of the fields of a given model are charged under the $U(1)$ gauge
groups, such that in general there is no distinct separation between the
observable and hidden sector gauge groups.  

In this class of models, the couplings are calculable
in string theory; techniques have been developed to
 calculate the superpotential~\cite{CFTO,C,worldsheet,chl,cceel2,cceel3}
in principle to all orders in the nonrenormalizable terms. 
 One generic feature of
the superpotentials  is that additional worldsheet
selection rules forbid terms otherwise allowed by gauge invariance.  However, 
the determination of the K\"{a}hler potential is more involved 
in part because the K\"{a}hler potential is not
protected by
supersymmetric non-renormalization theorems and
 thus receives corrections at all
orders in the string loop expansion.  Therefore, 
in the analysis that follows we
assume a minimal K\"{a}hler potential, for the sake of simplicity.

The analysis of this class of quasi-realistic models proceeds in several stages.
The first step is to address the presence of the anomalous $U(1)$ in the model.
The underlying superstring theory is anomaly-free, and hence there is a
standard mechanism in the four-dimensional effective theory in which the
axion-dilaton supermultiplet shifts under $U(1)_A$ in such a way that all
triangle anomalies are cancelled.  The anomaly cancellation mechanism
generates a nonzero Fayet-Iliopoulos (FI) contribution to the $D$- term of
the anomalous $U(1)$ at higher genus in string theory~\cite{DSW,ADS,AS}.
The FI term would break supersymmetry in the original
string vacuum, but certain scalar fields are triggered to acquire
large vacuum expectation values (VEV's) along $D$- and $F$- flat
directions~\cite{DSW,DK}.  The new ``restabilized" vacuum is
supersymmetric, with a gauge structure of reduced rank
(in particular, the anomalous $U(1)$ is broken), and a reduced number of
massless fields, as the fields which couple to the fields in the flat
direction can acquire string-scale masses and decouple from the theory.
Therefore, an analysis of the $D$- and $F$- flat directions is the
necessary first step in the investigation of the phenomenology of the
string model. (Issues of anomalous $U(1)$ in string models and string
(motivated) models were discussed in~\cite{genanomp,genanoms}.)

In a previous paper~\cite{cceel2}, we developed techniques to classify the
flat directions of a general perturbative heterotic superstring model with
an anomalous $U(1)$.  For the sake of simplicity, we chose to consider
flat directions formed of non-Abelian singlets only, and selected the
singlet fields with zero hypercharge to preserve the SM gauge group.  Our
method involves classifying the fields according to their anomalous charge
to see if flat directions that can cancel the FI term can be
formed. If such flat directions can be formed, we construct the superbasis
of all one-dimensional (i.e., that depend on one free VEV before imposing
the anomalous $D$ term constraint) $D$- flat directions under the
non-anomalous $U(1)$'s.  The elements of the superbasis with the
appropriate sign of the anomalous charge to cancel the FI term are the
building blocks of the $D_A$- flat directions of the model.  

For a subset of these $D$- flat directions,
the requirements of gauge invariance as well as a string calculation of the
superpotential to a given order suffice to prove $F$- flatness to all orders in
the nonrenormalizable superpotential.  Our method provides a systematic and
complete classification of the subset of the $D$- flat directions which can be
proven to be $F$- flat to all orders. Each of these flat directions
corresponds to a new restabilized string vacuum of a given model.  We
applied our method to a prototype string model, Model 5 of~\cite{chl}
(CHL5), in~\cite{cceel2}, and more recently
to a number of free fermionic string models in~\cite{cceel3}.  

The next stage of the analysis of this class of string models is to investigate
the implications of each flat direction of a given model.  In general, a
number of $U(1)$'s are broken in each flat direction, though we find that
usually at least one $U(1)$ in addition to $U(1)_Y$ remains unbroken.  The
couplings of the fields in the flat direction to the other fields in the
model lead to the generation of effective mass terms, such that some of
the fields acquire superheavy masses and decouple from the theory.
Effective trilinear couplings are also induced from
higher-dimensional terms for the remaining light states, with implications
for the phenomenology of the model. Effective nonrenormalizable terms
are also generated~\footnote{However, nonrenormalizable terms competitive
in strength are also present in the original superpotential, as well as
generated in a number of other ways, such as via the decoupling of heavy
states\cite{ceew}, a nonminimal K\"{a}hler potential, and the corrections
to the K\"{a}hler potential due to the large VEV's.}. We have developed
techniques to determine the effective bilinear and trilinear couplings to
all orders in the fields  with nonzero VEV's along a particular flat
direction. 
Once these terms are determined, they are exact to
all orders in the string genus expansion~\cite{witten}. The details of the
effective
superpotential strongly depend on the flat direction under consideration.

In this paper, we analyze a class of the flat directions 
obtained in~\cite{cceel2}
for the prototype string model (Model 5 in ~\cite{chl} (CHL5)), and
investigate the implications of the mass spectrum and effective trilinear
couplings.  We choose to consider flat directions with the maximal number
of fields, which also break the maximal number
of $U(1)$'s; in these directions, one $U(1)'$ is unbroken in addition to
$U(1)_Y$.
We generalize our techniques developed to prove $F$- flatness to  determine 
the effective renormalizable superpotential:
\begin{itemize}
\item {\bf Massless Spectrum.}
 For each flat direction, we
determine
the complete massless spectrum (at the string scale).
We find that
the flat directions considered share the undesirable feature that along
with the MSSM content there are additional massless exotics.
For the cases in which the $U(1)'$ is broken at the electroweak scale,
the exotic fermions remain light compared to the electroweak scale,
which is excluded by experiment.
This  feature indicates a general flaw of this type of
models.
\item {\bf Trilinear terms.}
Nevertheless, we proceed and  analyze the effective trilinear
couplings. For each flat direction we determine  all such couplings  (to
all
orders in the VEV's of fields in the flat direction)  in the observable
sector of the theory, i.e., the SM and $U(1)'$ sector of the theory. 
(In addition we include terms involving the hidden sector fields that  
play a role in the renormalization group analysis of the symmetry breaking scenarios, 
a topic of  a subsequent paper~\cite{cceelw2}.)
\end{itemize}

We then discuss the implications of these couplings. The predictions (from
$\alpha_s$) for the electroweak gauge couplings are presented.
At the level of the trilinear  superpotential,
we consider the fermion masses  and textures;  baryon- and lepton-number
violating couplings; $R$-parity breaking terms and the absence of a stable
LSP;   and the occurrence of
the non-canonical (``half'') $\mu$- term.  
We identify  types of symmetry breaking scenarios for  $U(1)'$, one
at the  electroweak scale~\cite{cl,SY,cdeel,lw,KM} and another one at an
intermediate scale~\cite{cl,cceel1}; and discuss the family
non-universality of $Z'$ couplings.

 While we calculated the mass spectrum and the effective trilinear terms
 (in the observable sector) 
 for all the flat directions (classified in~\cite{cceel2}),   in the paper
  we illustrate the techniques and present the detailed analysis along the
steps discussed above for {\it two representative flat directions}.

In the subsequent paper~\cite{cceelw2}  we plan to carry out further the
phenomenological consequences  by introducing  soft supersymmetry breaking
mass parameters; we shall analyze
specific  SM and $U(1)'$ symmetry breaking
patterns, consistent with experiment,  and the particle mass
spectrum at the electroweak scale.

The paper is structured as follows.  In Section \ref{II}, we review the
flat direction
analysis for this model given in \cite{cceel2}, and present our techniques
to compute the effective renormalizable superpotential for each flat
direction considered.
In  Sections~\ref{III} and ~\ref{IV}  the two respective 
 representative examples are
analysed in detail. Conclusions and a discussion of further
investigations are given in Section~\ref{V}.

\section{Flat Directions and Effective Couplings}
\label{II}
The model we choose to analyze is Model 5 of \cite{chl} (CHL5), 
which has the gauge group 
\begin{equation}
\{SU(3)_C\times SU(2)_L\}_{\rm obs}\times\{SU(4)_2\times SU(2)_2\}_{\rm hid}
\times U(1)_A\times U(1)^6.
\end{equation}
In addition to the MSSM fields, the particle content includes the additional
chiral superfields:
\begin{eqnarray}
&&6 (1,2,1,1) + (3,1,1,1) + (\bar{3},1,1,1) + \nonumber\\
&&4 (1,2,1,2) + 2 (1,1,4,1) + 10 (1,1,\bar{4},1) +\nonumber\\
&&8 (1,1,1,2) + 5 (1,1,4,2) + (1,1,\bar{4},2) +\nonumber\\
&& 8 (1,1,6,1) + 3 (1,1,1,3)+ 42 (1,1,1,1)\;\;,
\end{eqnarray}
where the representation under $(SU(3)_C,SU(2)_L,
SU(4)_2,SU(2)_2)$ is indicated.  
The complete list of fields with their $U(1)$ charges are presented in Tables
Ia-Ic. $Q$, $u$, and $d$ denote quark  candidates (doublets or singlets),
with ${\cal D}$ reserved for an exotic quark singlet.
$h$ generically denote Higgs or lepton doublet candidates, $e$
represents  possible
lepton singlet
candidates, while $\varphi$ is left for other singlets. Capital letters
are reserved for fields in non-trivial representations of the hidden
sector non-Abelian groups.
 
The SM hypercharge is determined as a linear combination of the non-anomalous
$U(1)$'s, subject to basic phenomenological criteria 
(see \cite{cceel3} for more
details). We require three families
of quarks and leptons, as well as at least two
candidate electroweak Higgs doublets, with conventional hypercharges.  We also
require  grouping of all fields  which are charged under $U(1)_{em}$ 
and/or transform under  $SU(3)_C$ 
into mirror pairs;  this is a prerequisite to  have a possibility of 
avoiding  
 exactly massless charged  and/or colored particles
 in
the theory.  In this model, these criteria lead to the unique hypercharge
definition (first presented in \cite{chl}):
\be
\label{hyp}
Y=\frac{1}{96}(-8Q_2-3Q_3-8Q_4-Q_5+Q_6),
\ee
[normalized to give $Y$(quark doublet)$=1/6$], with Ka\v c-Moody level 
$k_Y=\frac{11}{3}$.(We calculate $k_Y$ using the universal Green-Schwarz
relations; for more details, see \cite{cceel3}.)

\subsection{Analysis of Flat Directions}

The presence of the anomalous $U(1)$ leads to the generation of a nonzero
Fayet-Iliopoulos (FI) term $\xi$ in the corresponding $D$- term at
genus-one~\cite{DSW,ADS} (at
genus two for the  dilaton tadpole~\cite{AS}) in string 
theory with 
\begin{equation}
\xi = \frac{g^2_{\rm string}M_P^2}{192\pi^2}\Tr Q_A\, ,
\label{fid}
\end{equation}
in which $g_{\rm string}$ is related to the  gauge coupling $g$
by the relation $g_{\rm string}=g/\sqrt{2}$~\cite{Kap} 
  ($g$ is  normalized according to the standard (GUT)  conventions,
 i.e., ${\rm Tr}T_aT_b=\delta_{ab}/2$ for the generators of the
fundamental representation of $SU(N)$) and $M_P$ is the reduced Planck
mass, such that $M_P = M_{
Pl}/\sqrt{8\pi}$, with $M_{ Pl} \sim 1.2\times 10^{19}$ GeV.

The FI $D$- term triggers the scalar components $\varphi_i$ of certain chiral
superfields $\Phi_i$ to acquire VEV's in such a way
that $D$- and $F$- flatness conditions are satisfied~\cite{DSW,DK}.  
In principle, the fields 
which acquire VEV's may be those with nontrivial representations under the
observable sector or the non-Abelian hidden sector.  For simplicity, we 
restrict our consideration to the non-Abelian singlet fields of the model, and
select only those fields with zero hypercharge to preserve the SM gauge group at
the string scale. The $D$- and $F$- flatness conditions for these fields are given
by 
\begin{eqnarray}
D_{\rm A} &=& \sum_i Q^{(A)}_i |\varphi_{i}|^2 + \xi
= 0\, \\
\label{anomd}
D_{a} &=& \sum_{i} Q^{(a)}_{i}|\varphi_{i}|^2 = 0\\
\label{nanomd}
F_{i} &=& \frac{\partial W}{\partial \Phi_{i}} = 0; \,\, W  =0.
\label{ff}
\end{eqnarray}
We list the non-Abelian singlets of the model with their $U(1)$ charges in Table
I.  

In a previous paper \cite{cceel2}, we presented techniques to classify the $D$-
and $F$- flat directions of a general
 perturbative heterotic string model with an
anomalous $U(1)$, and illustrated the method by applying it to this model.  We
summarize the method and repeat the conclusions here for the sake of
completeness, and refer the reader to \cite{cceel2} for more details.  

First, the $D$- flat directions associated with the non-anomalous $U(1)$'s are
determined, by making use of the one-to-one correspondence of $D$- flat
directions with holomorphic gauge-invariant monomials (HIM's) of  chiral
superfields~\cite{flat1,flat2,flat3}, constructed 
from the non-Abelian singlets. We construct the set of
all one-dimensional HIM's, which we refer to as the superbasis.  
The complete set
of $D$-flat directions under the non-anomalous $U(1)$'s can
 be obtained by
multiplying the elements of the superbasis. 
 The elements of the superbasis with
anomalous charge opposite in sign to the FI term $\xi$ are flat with
respect to the $D$- term of the anomalous $U(1)$.  This subset of 
the superbasis
elements, which we denote as $\{P_{\alpha}\}$, 
are the building blocks for the
$D$- flat directions of the model.

We presented the superbasis in \cite{cceel2}, and
showed that there are five such classes of $D_A$- flat elements:
\bear
\label{pis}
P_1&=&\langle \varphi_{28} ,\varphi_{27}^2\rangle\nonumber,\\
P_2&=&\langle
\varphi_{4},\varphi_{10},\varphi_{30},\varphi_{27}^2\rangle\nonumber,\\
P_3&=&\langle
\varphi_{12},\varphi_{2},\varphi_{30},\varphi_{27}^2\rangle\nonumber,\\
P_4&=&\langle
\varphi_{4},\varphi_{2},\varphi_{16},\varphi_{30},\varphi_{27}^2\rangle\nonumber,\\
P_5&=&\langle
\varphi_{12},\varphi_{10},\varphi_1,\varphi_{30},\varphi_{27}^2\rangle.
\eear
To these, one should add similar monomials obtained by replacing
some field by its copy ($\varphi_{2}\rightarrow \varphi_{3}$,
$\varphi_{4}\rightarrow \varphi_{5}$, $\varphi_{10}\rightarrow \varphi_{11}$,
$\varphi_{12}\rightarrow \varphi_{13}$, $\varphi_{28}\rightarrow \varphi_{29}$).   
Therefore, every $D_A$- flat direction can be
obtained from the set $\{P_{\alpha}\}$ by 
\be
P=P_\alpha N,
\ee
with $N$ some HIM (not necessarily with $Q_A>0$).

To address the $F$- flatness of the $D$- flat directions, we note that there are
two classes of terms in the superpotential that can lift a general $D$- flat
direction $P$.  First, there can be terms which are formed only of the fields in
the flat direction $P$:   
\be
\label{wa}
W_A\sim\left(\Pi_{i\in P} \Phi_i\right)^n,
\ee
in which the coefficients (which depend on inverse powers of $M_{Pl}$) are
not displayed explicitly. The flat direction $P$
will be said to be type-A if such an invariant is 
allowed by gauge symmetries. If such an invariant exists,
 there are an infinite number
of terms which can lift the flat direction, because this
invariant can appear to any power in the superpotential. 
The type-A directions will remain $F$- flat only if  string selection rules
(e.g. $R$ parities) conspire to forbid the infinite number of $W_A$ terms,
which is 
difficult to prove
in general.  

The other class of terms are of the form
\be
\label{wb}
W_B\sim\Psi\left(\Pi_{i\in P} \Phi_i\right),
\ee
with  $\Psi\notin P$. 
A flat direction will be denoted as type-B if gauge 
invariance only allows $W_B$
terms.  In contrast to the case with the $W_A$ terms, gauge invariance constrains
the number of $W_B$ terms which can exist to 
a finite number. By doing a string calculation of the
superpotential to a finite order, the presence of these terms can be checked
explicitly, and if such terms are absent the flat direction is proved to be $F$-
flat to all orders in the nonrenormalizable superpotential.
We take a conservative approach by restricting our consideration to the type-B
directions which can be proved to be $F$- flat to all orders.  In doing so,
 we are
of course neglecting certain type-A directions that may be $F$- flat due to
``string  selection-rules'' of the model.

It is straightforward to show~\cite{cceel2} that type-B directions are
formed only from carefully combining the $P_{\alpha}$'s.  In
this model, we found that $P_4$ and $P_5$ are not $F$- flat, and thus the $F$-
flat directions are formed from $P_1$, $P_2$ and $P_3$ (and the primed versions of
them involving copies of the fields).  We present the complete
 list of $F$- flat directions
in Table II.  

This table demonstrates that there are a range of flat directions, which break
different numbers of the non-anomalous $U(1)$'s.  We choose to analyze the
$P_2P_3$ and $P_1P_2P_3$ flat directions, which
break the maximum number of $U(1)$'s; these
directions all leave an additional $U(1)'$ as well as $U(1)_Y$ unbroken.
The unbroken $U(1)'$ is given by
\be
Y'=\frac{1}{1200}(-130Q_2-14Q_3+148Q_4-51Q_5+51Q_6),
\label{u1p}
\ee
with $k_{Y'} = 4167/250 \simeq 16.67
$.

The $D$- term constraints for the VEV's of the fields in
the most general $P_1P_2P_3$ direction yield the relations
\be
\label{vevs}
\begin{array}{lcl}
|\varphi_{27}|^2=2x^2, &\;\;\;\;&
|\varphi_{28\,(29)}|^2=x^2-|\psi_1|^2,\nonumber\\
|\varphi_{30}|^2=|\psi_1|^2, &\;\;\;\; &
|\varphi_{4\,(5)}|^2=|\psi_2|^2,\nonumber\\
|\varphi_{2\,(3)}|^2=|\psi_1|^2-|\psi_2|^2, &\;\;\;\; &
|\varphi_{12\,(13)}|^2=|\psi_1|^2-|\psi_2|^2,\nonumber\\
|\varphi_{10\,(11)}|^2=|\psi_2|^2,&\;\;\;\;&
\end{array}
\ee
with 
\be
\label{xdef}
x^2=-\frac{\xi}{64}.
\ee
In this model $\Tr Q_A=-1536$, and $x=0.013M_{Pl}$.
$|\psi_{1,2}|$ are free VEV's of the moduli space, subject to the restrictions
that 
\be
\label{restrict}
x^2\ge |\psi_1|^2\ge |\psi_2|^2,
\ee 
to ensure the positivity of the VEV-squares.
Simpler flat directions are recovered by setting the free VEV's to particular
values; for example, setting
$|\psi_1|^2=|\psi_2|^2=0$ yields the solution for 
the VEV's of $P_1$.  Therefore, there can be an enhanced number of $U(1)$'s at
particular points in the moduli space. The $P_2P_3$ directions are obtained by
setting $|\psi_1|^2=x^2$, which gives $|\varphi_{28\,(29)}|^2=0$.  

In some cases, a judicious choice of the 
copies of the fields
allows for $F$- flatness without imposing any constraints on the free VEV's.
However, other possible flat directions can be formed by imposing constraints on
the free VEV's in such a way as to cancel contributions from different $F$- terms,
i.e., of the type $\varphi_9(\varphi_{4}\varphi_{10\, (11)}+
\varphi_{2\,(3)}\varphi_{12})$~\cite{cceel2}.
For example, the directions denoted by $|_F$ are obtained by imposing
$|\psi_1|^2=2|\psi_2|^2$, and  $\pi$  phase difference  between 
VEV's  of  the   $\varphi_{4}\varphi_{10\,(11)}$ and 
$\varphi_{2\,(3)}\varphi_{12}$ terms~\cite{cceel2}~\footnote{Throughout
the paper we
assume that these VEV's are real. For the model discussed 
the introduction of complex 
phases for  these VEV's can be
absorbed into the redefinition of the  remaining 
fields in the effective  superpotential, and thus it
does not affect the physics at the level of the effective  bilinear and 
trilinear terms.}.  The complete list of all
such flat directions is given in Table II.

\subsection{Mass Spectrum}
For each flat direction, effective mass terms
for fields $\Psi_i$, $\Psi_j$ ($\Psi_{i,j}\notin  \{\Phi_k\}$)
 may be generated by the coupling of these
fields to the fields $\Phi_i$ in the flat direction, such that
\begin{equation}
\label{effmass}
W\sim\Psi_i\Psi_j\left(\Pi_{k\in P}\Phi_k\right)\ .
\end{equation}
The fields with effective mass terms will acquire $F$-term string-scale
masses and decouple from the low-energy theory. 
In cases in which $F$-flatness occurs via cancellations, other fields 
coupled linearly to the flat directions fields also get 
heavy masses (we will later study such a case). 

In addition to these large masses induced by $F$-terms, $D$-terms can also 
make some (combination) of the fields related to the flat direction heavy
(all other fields do not feel the presence of large VEVs in the $D$-terms
because of the $D$-flatness conditions). 
Particular combinations of the real components of the fields entering the 
flat direction gain a mass of order $g\sqrt{\xi}$ (through 
$D$- terms) and become degenerate with the massive $U(1)$ gauge bosons,
completing,  along with a Dirac 
fermion (a neutralino), a massive vector
multiplet.  This is guaranteed by the fact that supersymmetry remains
unbroken in the restabilized vacuum, so that the spectrum must arrange
itself in supersymmetric multiplets\footnote{
In more detail this works as follows. Using the matrix
$m_{ai}=\sqrt{2}g_aQ_i^{(a)}\langle\varphi_i\rangle$, the squared-mass
matrix of $U(1)$ gauge bosons is $mm^T$ while that of real scalars,
coming from $D$-terms is $m^Tm$. It is a simple exercise to show that
the non-zero eigenvalues of $mm^T$ and $m^Tm$ are equal and in exact
one-to-one correspondence. The presence of non-zero $F$-term scalar masses
does not spoil this correspondence when $V=0$: they simply give mass to
zero-eigenvalues of $m^Tm$.}. 
The imaginary parts of these fields become the longitudinal components
of the massive $U(1)$ gauge bosons. 
(This scenario  thus exhibits all the features of
the Higgs mechanism in  $N=1$ supersymmetric theory.)

For  directions which are flat due to cancellations of $F$- term
contributions, some  (complex) fields in the flat
direction get masses of order ``[Yukawa] $\times$ [field VEV]'' 
and form, along with
its superpartner, a massive chiral superfield. (As mentioned above, the
fields with zero
VEV's which couple  linearly
in these terms also acquire mass of the same order.)  Remaining fields in
the flat direction,
including the real parts of the fields whose VEV is  not fixed
by the flatness conditions, as well as imaginary parts not removed by the Higgs
mechanism, will stay massless and will appear in the low energy
theory\footnote{This is intuitively clear: (scalar) field excitations
along the flat direction are massless while excitations transverse to
the flat direction are massive. In the presence of a FI term
however, the flat direction can be reduced to a single point, with all 
scalar field excitations massive.} as massless chiral superfields (moduli).

For the flat directions of the type $P_2P_3|_F$, which has all the VEV's fixed,
  all these  (six) fields are  heavy.
 Namely, the Higgs mechanism ensures that five imaginary components   are
 Goldstone bosons giving  mass to the  gauge bosons of five (including
 anomalous) $U(1)$'s, and the accompanying five real components get mass,
thus  completing  five massive vector supermultiplets.
The remaining complex field  gets a mass from the superpotential terms
of the type
$\varphi_9(\varphi_{4}\varphi_{10\,(11)}+
\varphi_{2\,(3)}\varphi_{12})$, i.e.,
the terms  which require  additional constraints on the VEV's in
order to
 ensure $F$- flatness.
Note that due to these terms  $\varphi_9$  also acquires a mass. 

For the
flat directions of the type  $P_1P_2P_3$, there  are two
free VEV  parameters, since now 
an additional field participates in the vacuum restabilization and there are no
 constraints on the VEV's from $F$- flatness constraints (as opposed to the 
$P_1P_2|_F$ flat directions), and the
corresponding analysis  shows that now there are 
two  massless complex fields, which act as  moduli in the space of
restabilized vacua.

\subsection{Effective Couplings}

The spectrum of the low-energy theory not only arranges itself in SUSY
multiplets, but also, the interactions of the light particles can be
described by an effective superpotential which contains only the light
degrees of freedom. (The moduli fields associated
with the flat
direction in question are absent from this superpotential.) 
In addition to the trilinear couplings of the original superpotential, 
effective renormalizable interactions for the light fields may also be
generated via
\begin{equation}
\label{efftril}
W\sim\Psi_i\Psi_j\Psi_k\left(\Pi_{l\in P} \Phi_l\right)\ .
\end{equation}  
In principle, effective nonrenormalizable terms at each order will also be
generated from higher-order nonrenormalizable terms 
in the superpotential in this
way.  However,
there are many other sources for effective
 nonrenormalizable terms (such as via 
the decoupling of the heavy fields~\cite{ceew}) which 
are competitive in strength.
A complete classification of the effective nonrenormalizable terms is
beyond the scope of this paper.

The method for determining the effective mass terms and trilinear interactions
for each flat direction is
similar to the strategy for determining the $W_B$ terms in the superpotential
when testing for $F$- flatness.
First, we construct all the bilinear and trilinear terms which are gauge
invariant under the unbroken gauge group of the model after the vacuum
restabilization (for the $P_2P_3$ and $P_1P_2P_3$ directions, this
includes $U(1)_Y$ and $U(1)'$ as well as the non-Abelian gauge
groups). We then treat each term as a composite field, 
and construct all possible
terms that are gauge invariant under all of the $U(1)$'s in 
the theory (including
the anomalous $U(1)$) which involve the fields in the flat direction and are
linear in the composite field.  The next step is 
to calculate explicitly whether
each gauge invariant term is present in the superpotential,
 or forbidden by string
selection rules.  In practice, the requirements of 
gauge invariance give the order
to which the superpotential must be calculated to determine the full effective
renormalizable superpotential.

Once the mass terms are determined, it is straightforward to determine the
complete mass spectrum of the model.  The trilinear interaction terms are then
written in terms of the mass eigenstates, so that the decoupling theorem can be
applied to the terms involving the superheavy fields.

In  free fermionic constructions the elementary trilinear
superpotential terms have coupling strengths ${\cal O}(g)$: the
typical value  is given  by $\sqrt{2}g_{\rm string}=g$,  where again
  $g$ is the
gauge coupling.    However, the
introduction of the ``Ising worldsheet fields'' in  more involved
constructions (e.g., in ~\cite{chl}), allows  also for Yukawa couplings
$g/\sqrt{2}$ and $g/2$~\cite{worldsheet}. 
In general, the coefficients $(\equiv\alpha_{K+3}/M_{Pl}^K)$  ($K>0$) of
 the nonrenormalizable superpotential terms of order
$K+3$ are given by the relation: 
\begin{eqnarray}
\frac{\alpha_{K+3}}{M_{Pl}^K}&=&g_{\rm 
string}\left(\frac{g_{\rm 
string}}{2\pi}\right)^{K}(\sqrt{2\alpha'})^KC_{K}I_{K}\nonumber\\
&=&g_{\rm string} 
\left(\sqrt{\frac{8}{\pi}}\right)^{K}\frac{C_{K}I_{K}}{M^{K}_{Pl}}\ ,
\end{eqnarray}
where $\alpha'=16\pi^2/(g_{string}^2M_{Pl})$ is the inverse string tension,
 $C_{K}$ is a
coefficient of ${\cal O}(1)$ which includes different renormalization factors 
in the operator product expansion (OPE) of the string vertex operators
(including the target space gauge group Clebsch-Gordan coefficients),
and $I_{K}$ is a world-sheet integral.  The values of $I_1$ and $I_2$ have been
computed numerically by several authors~\cite{worldsheet} with the  typical
values   that are in the range 
$I_1\sim70$, $I_2\sim400$. (See~\cite{CEW}
where special attention is paid to restoring the correct factors and units.)  

The coefficients of the effective trilinear terms are then given by  
$\sim (\alpha_{K+3}/M_{Pl}^K)|\varphi_i|^K$.  Using $x$ as the
typical scale of the VEV's, we find that the terms from the fourth order have
effective Yukawa coupling strengths $\sim 0.8C_1$, while the fifth order terms
have coupling strengths $\sim 0.1C_2$.  (We took $\sqrt{2}g_{string}=g\sim 0.8$,
since this is the value typically obtained for the model discussed.)  
 Therefore, compared to the typical elementary trilinear term 
 $\sqrt{2}g_{string}=g\sim 0.8$ the fourth order terms are 
competitive in strength to the trilinear terms
 of the original superpotential, while
the higher order contributions are suppressed. 
 Of course, the precise values for
each term will depend on the particular fields involved. In particular, the
coupling strengths can depend on the free VEV's of the flat direction, 
and hence
are parameters that can be varied in the analysis of the model.  

The structure of the effective trilinear
 couplings strongly depends on the flat
direction under consideration. 
We have determined the  mass spectrum and the  effective trilinear
superpotential terms (in the observable sector) for  the directions classified in
Table II~\cite{cceel1}. In the following, we 
 analyze  the details of the  two flat
directions, $P_1'P_2'P_3'$ and $P_2P_3|_F$, as representative examples.   These
directions encompass  general features of the whole class of flat directions
(Table II) and demonstrate the nature of the massless spectrum and its
phenomenological implications. The first  flat direction is ``minimal''
in the sense that there are a minimal number of 
 surviving trilinear couplings  of  the massless observable sector. The second
one has a  richer structure of such couplings, with implications for, e.g., the
fermion texture, and baryon and lepton violating processes.

\section{$P_1'P_2'P_3'$ Flat Direction}
\label{III}
\subsection{Effective Superpotential}

This flat direction involves the set of fields  
$P_1'P_2'P_3'=\{\varphi_{2},\varphi_{5},\varphi_{10},\varphi_{13}, 
\varphi_{27},\varphi_{29},\varphi_{30}\}$ (see Table~II).   The VEV's for
the fields correspond to the most general case given in (\ref{vevs}), such
that they depend on two free parameters (that are constrained to be
bounded from above by the value of the FI term as dictated by eq.
(\ref{restrict})).

The effective mass terms are computed for this flat direction using the
techniques described in the previous section, with the
result
~\footnote{The effective couplings in this and the subsequent section involve  
 third and fourth order couplings that are  
modified from those quoted in \cite{chl}. This Modification is due to two
effects: (i) the correctly implemented picture changing procedure 
in the calculation of couplings  introduces a number of
additional couplings at the fourth order, and (ii) the  implementation of
the tests calculating contributions to the  correlation functions from
the real left-moving worldsheet fermions excludes a number of couplings
involving some of the non-Abelian hidden sector fields.
The full superpotential up to the fifth order will be presented
elsewhere.}: 


\begin{eqnarray}
\label{p123massw}
W_{M}&=&gh_f\bar{h}_b \langle \varphi_{27}\rangle+gh_g\bar{h}_d \langle
\varphi_{29}\rangle+
\frac{\alpha^{(1)}_{4}}{M_{Pl}}h_b\bar{h}_b \langle
\varphi_{5}\varphi_{10}\rangle + 
\frac{\alpha^{(2)}_{4}}{M_{Pl}}h_b\bar{h}_b \langle
\varphi_{2}\varphi_{13}\rangle
\nonumber\\
&+&{g\over {\sqrt{2}}}(e^c_de_b+e^c_ge_a)\langle 
\varphi_{30}\rangle +
{g\over {\sqrt{2}}}(\varphi_1 \varphi_{15}+\varphi_{4}\varphi_{9})\langle
\varphi_{10}\rangle\nonumber\\ &+&
{g\over {\sqrt{2}}}(\varphi_7 \varphi_{16}+\varphi_{9}\varphi_{12})\langle
\varphi_{2}\rangle 
+
{g\over {\sqrt{2}}}(\varphi_6 \varphi_{26}+\varphi_{8}\varphi_{23}+\varphi_{14}\varphi_{17})\langle
\varphi_{29}\rangle+\frac{\alpha^{(3)}_{4}}{M_{Pl}}\varphi_{21}\varphi_{25}\langle
\varphi_{27}\varphi_{29}\rangle\nonumber\\ &+&
{g\over {\sqrt{2}}}(\bar{F}_1 F_1 +\bar{F}_2 F_2)\langle \varphi_{30}\rangle
+{g\over {\sqrt{2}}}S_3S_5 \langle \varphi_{5}\rangle+
{g\over {\sqrt{2}}}S_1S_5 \langle
\varphi_{13}\rangle \, .
\end{eqnarray}
The coefficients of the elementary trilinear
terms, 
 equal to $g$  or $g/\sqrt{2}$,  are  displayed explicitly. 
It is straightforward to determine the mass eigenstates, and we list the massive
and massless states in Table III.
(The mass spectrum of fields with non-zero VEV's in the flat directions
were discussed in Section IIA and are not explicitly displayed in the
Tables.)

We then determine the effective trilinear couplings involving
the observable sector
fields. In addition, we  
inspect the effective trilinear self-couplings of the non-Abelian singlets 
and the effective trilinear couplings 
 of non-Abelian singlets to the hidden sector fields which could affect the RGE
 for the couplings in the observable sector. 
The result is the following: 
\begin{eqnarray}
W_{3}&=&gQ_cu^c_c\bar{h}_c+gQ_cd^c_bh_c
+\frac{\alpha^{(4)}_{4}}{M_{Pl}}Q_cd_d^ch_a\langle \varphi_{29} \rangle
+{g\over {\sqrt{2}}}e^c_ah_ah_c+
{g\over {\sqrt{2}}}e^c_fh_dh_c  \nonumber\\&+& 
\frac{\alpha^{(1)}_{5}}{M^2_{Pl}}e^c_hh_eh_a\langle
\varphi_{5}\varphi_{27}\rangle+
\frac{\alpha^{(2)}_{5}}{M^2_{Pl}}e^c_eh_eh_a\langle
\varphi_{13}\varphi_{27}\rangle+gh_f\bar{h}_c\varphi_{22}+
gh_b\bar{h}_c\varphi_{20} \nonumber \\
&+& \frac{\alpha^{(5)}_{4}}{M_{Pl}}S_2 S_6 \varphi_{20} \langle \varphi_{27}\rangle
+ \frac{\alpha^{(3)}_{5}}{M^2_{Pl}}S_2 S_6 \varphi_{22} \langle \varphi_{2} \varphi_{13}\rangle + \frac{\alpha^{(4)}_{5}}{M^2_{Pl}}S_2 S_6 \varphi_{22} \langle \varphi_{5} \varphi_{10}\rangle 
%
\,.
\end{eqnarray}
The result is expressed in terms of the mass eigenstates and the VEV's, and 
the decoupling theorem is applied to the terms involving the heavy fields:
\begin{eqnarray}
W_{3}&=&gQ_cu^c_c\bar{h}_c+gQ_cd^c_bh_c+{g\over {\sqrt{2}}}e^c_ah_ah_c+
{g\over {\sqrt{2}}}e^c_fh_dh_c+
\frac{\sqrt{2} \alpha^{(1)}_{5}x^2}{M_{Pl}^2}\lambda_2
e^c_h h_eh_a\nonumber\\&+&
\frac{\sqrt{2} \alpha^{(2)}_{5}x^2}{M_{Pl}^2}
\sqrt{\lambda_1^2-\lambda_2^2} \
e^c_eh_eh_a +g\bar{h}_ch_b'\varphi_{20}' 
+\frac{\alpha^{(4)}_4 x}{M_{Pl}}\sqrt{1-\lambda_1^2}Q_cd_d^ch_a \nonumber\\
&+&\frac{\sqrt{2}\alpha^{(5)}_{4}x}{M_{Pl}} \frac{1}{\sqrt{1+r^2}}(\varphi_{20}^{'}+r\varphi_{22}^{'})S_2 S_6 \nonumber \\
&+& \frac{x^2 [\alpha^{(3)}_{5}(1-\lambda_{1}^{2})+\alpha^{(4)}_{5} \lambda_{2}^{2}]} {M^2_{Pl}}\frac{1}{\sqrt{1+r^2}}(-r \varphi_{20}^{'}+\varphi_{22}^{'})S_2 S_6 ,
\end{eqnarray}
in which 
\begin{equation}
\lambda_2\equiv\frac{|\psi_2|}{x} \le \lambda_1\equiv\frac{|\psi_1|}{x}
\le 1,
\end{equation}
$\varphi_{20}'=\frac{1}{\sqrt{1+r^2}}
(\varphi_{20}-r\varphi_{22})$ and $\varphi_{22}'=\frac{1}{\sqrt{1+r^2}}
(r \varphi_{20}+\varphi_{22})$ , with
$r \equiv [\alpha_4^{(1)}
\lambda_2^2+\alpha_4^{(2)}(\lambda_1^2-\lambda_2^2)]x /(\sqrt{2}g M_{Pl})$.
$h_{b}^{'}$ is defined in Table IIIb. In the numerical analysis, 
$\lambda_1, \lambda_2$ are parameters that can be varied.

\subsection{Implications}

The effective superpotential has a number of interesting implications:
\begin{itemize}

\item {\bf Massless states:}

There are a large number of states that remain massless, as indicated in
Table IIIb.   These states include both the usual  MSSM states and  
related exotic (non-chiral under $SU(2)_L$) states,
such as a fourth ($SU(2)_L$ singlet) down-type quark,
extra fields with the same quantum numbers as the lepton singlet superfields,
and extra Higgs doublets.
There are other massless states with exotic quantum numbers (including
fractional electric charge), and states which are non-Abelian representations
under both the hidden and observable sector gauge groups and thus directly mix
the two sectors.  
The scalar components of these superfields may acquire masses via soft
supersymmetry breaking.
 However, within our set of assumptions there is no mechanism to give many of
the fermions significant masses~\footnote{One possible mechanism is to invoke
a non-minimal
K\"{a}hler potential. Another one, which is not possible for this
particular flat direction, is to utilize an intermediate scale, 
as discussed in the next Section.}.
As discussed earlier, there are additional massless states
(moduli) associated
with the fields which appear
in the flat direction but which are not fixed.
These are not listed in Table IIIb. 

\item {\bf $U(1)'$ charges of light fields:}

To invoke an intermediate scale $U(1)'$ symmetry breaking scenario (as
discussed in \cite{cceel1}), which can lead to a mechanism to give significant masses to the
additional light fields via higher-dimension operators, it is necessary to have at
least one pair of $Y=0$ singlets which remain massless after vacuum
restabilization
which have $U(1)'$ charges opposite in sign (to allow for the breaking to occur
along a $D$- flat direction). An inspection of Table Ic indicates that the
singlet field $\varphi_{25}$ is required for this scenario; however, this field
acquires a string-scale mass for this direction, and decouples from the theory.  We
conclude that in this case, an intermediate scale scenario is not possible, and
hence the breaking of the $U(1)'$ is necessarily at the electroweak
scale. (We do not consider more complicated scenarios in which
the $U(1)'$ could be broken along with some of the hidden non-Abelian
groups.) 

As discussed in \cite{cl,cdeel,lw}, several
scenarios exist which can lead to the
possibility of a realistic $Z-Z'$ hierarchy. 
The scenario in which only the two
MSSM Higgs fields $h_c$, $\bar{h}_c$ acquire VEV's breaks both $U(1)_Y$
and $U(1)'$
(because the $U(1)$ charges of these fields are not equal and opposite),
but leads to a $Z'$ which is ${\cal O}(M_Z)$, which is already excluded.  
The scenario in which the symmetry breaking is driven by a large
trilinear
coupling (described
in \cite{cdeel}) is also not feasible because the $U(1)'$ charges of the
relevant Higgs fields (which have $Y=\pm 1/2$) are opposite in sign, and
thus do not allow for a
small mixing angle
Therefore, the only remaining possibility is to have a scenario in which
the
symmetry breaking is characterized by a large (${\cal O}({\rm TeV})$)
SM singlet VEV ($\langle \varphi_{20}^{'} \rangle$), with the $SU(2)_L \times U(1)_Y$ breaking at a lower
scale due to accidental cancellations.

In addition, the $U(1)'$ charges of the observable sector fields
indicate that the
$Z'$ couplings are family nonuniversal. In the quark sector, the largest
couplings are to the third family, with smaller (equal) couplings to
the first two. There is a large coupling to the exotic
${\cal D}_a$. In the lepton sector, where the family assignments are less 
clear, there are unequal couplings to all families. Family mixing between
the quarks or between the leptons would lead
to flavor changing neutral current (FCNC)
effects in the $Z'$ and (through
$Z-Z'$ mixing) $Z$ couplings, while mixing with exotics would
induce FCNC for both the $Z$ and $Z'$ directly.
(Of course, the present model does not
have a satisfactory way to introduce such family mixings.)

\item {\bf $L$- violation:}


The doublets $h_a, h_{b}', h_c, h_d$, and $h_e$ can in principle
(i.e., before examining their superpotential couplings), be identified
as either Higgs or lepton doublets.
$h_{c}$,
as well as $\bar{h}_{c}$, should clearly be
identified as Higgs doublets from their couplings to ordinary quarks.
We will also identify $h_{b}'$ as a Higgs doublet, so that the
$\bar{h}_ch_b'\varphi_{20}'$
term conserves lepton number.
The remaining $h_a$, $h_d$ and $h_e$ are candidates for lepton doublets.
In particular, the couplings of $h_a$ and $h_d$ to the Higgs doublet $h_c$
indicate that these are the doublets corresponding to the $\mu$ and $\tau$
leptons.
There is no difficulty with $h_d$. However, the leptoquark
coupling $Q_c d_d^ch_a$ (in which $Q_{c}=(t, b)^{T}$ and $d_d^c$ can be either $d^c$, $s^c$, or
the exotic quark $D^c$) as well as the $e^c_hh_eh_a $ and
$e^c_eh_eh_a $ terms would then violate lepton number by one
unit, and similarly there is no conserved $R$ parity in this model.
(The strength of the $L$- violating coupling $Q_c d_d^ch_a$, coming
from the 4th order, could be reduced by choosing the free parameter
$\lambda_1$ close to 1, while allowing the relevant massive fields to still
have string scale masses.)


One consequence of these couplings is that there is no
stable LSP in this model. For example, a neutral gaugino could decay
into the fermion $e^c_h$ and its (virtual) scalar conjugate $\tilde{e}_h$,
followed by $\tilde{e}_h \rightarrow e^-_e \nu_a$ or $e^-_a \nu_e$, where
we use the notation $h_{a,e}=(\nu, e^-)_{a,e}^T$ (See Figure 1a).  
In principle, Majorana neutrino mass terms can also be generated at one
loop from the $e^c_{h,e}h_eh_a$ or $Q_c d_d^c h_a$ coupling. However,
such terms are absent in this model because a nonzero VEV for $h_e$ or
$h_a$ is required \footnote{For a discussion of neutrino masses in
models with $R$ parity violation, see \cite{majoranamass}.}.


\begin{figure}
\centerline{
\hbox{
\epsfxsize=6.0truein
\epsfbox[70 32 545 740]{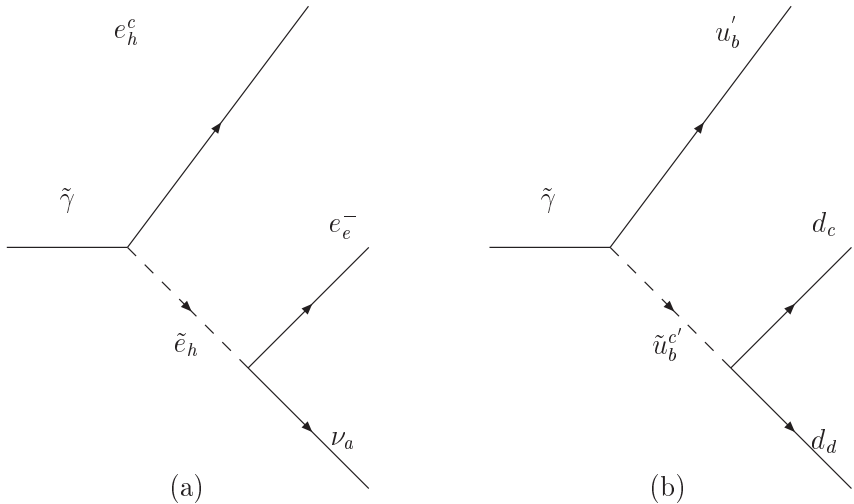}
}
}
\vskip -5.7 truein
\caption{Diagram for gaugino decay into (a) three leptons ($\Delta
L=1$), which occurs for the $P_{1}^{'}P_{2}^{'}P_{3}^{'}$ direction;
(There are similar diagrams for decay into $t d_d^c e_a^-$ or $b d_d^c
\nu_a$.) (b) three quarks ($\Delta B=1$), which occurs for the 
$P_{2}P_{3}|_{F}$ direction. $u_b^{'}$, $d_c$ and $d_d$ are $SU(2)$
singlets (i.e., the conjugates of $u_b^{c'}$, $d_c^c$ and $d_d^c$).   }
\end{figure}

\item {\bf Effective Yukawa couplings:}

The superpotential not only does not have an elementary $\mu$ term, but also
does not have the usual effective $\mu$ term \cite{cl,SY,cdeel} 
of the form $\bar{h}_c h_c \varphi$ for any SM singlet
$\varphi$.
Gauge
invariance and string selection rules forbid the presence of
this effective $\mu$ term for all of the flat directions considered for
this model.
There is, however, a  non-canonical 
$\mu$ term
$\bar{h}_ch_b'\varphi_{20}'$, which couples to only one of the
 the ordinary Higgs doublets ($h_{c}$ and ${\bar h}_c$). 
In a subsequent paper~\cite{cceelw2} we shall 
analyze the running of the Yukawa couplings and  an acceptable gauge
symmetry breaking pattern that can be obtained even without the soft
breaking
``$B \mu$'' found in the MSSM. However, the lack of a 
$\mu$ term  of the form ${\bar h}_a h_c \varphi$ leads to an 
unwanted massless chargino and neutralino, while the absence of this or of a canonical 
$\mu$ term  of the type ${\bar h}_c h_c \varphi$ leads to a second almost
massless neutralino and an unwanted approximate global $U(1)$ symmetry.

In addition, the superpotential for this direction has the feature that only the
third quark family has large
Yukawa couplings, which is a desirable feature.  However,
the Yukawa couplings also indicate $t-b$  and $\tau -\mu$ Yukawa 
unification with  the equal  string scale Yukawa couplings $g$ and
$g/\sqrt{2}$, respectively~\footnote{This type of Yukawa coupling
unification is  stringy in nature and different from the standard GUT
considerations. The 
obtained hierarchy between the  lepton and quark Yukawa couplings is due to the
fact that the  ``canonical'' candidates for the lepton doublets (which would have
had the same Yukawa couplings as quarks) are massive for the  flat
directions considered, and thus the lepton Yukawa couplings involve  fields
that  are usually identified with additional copies of the (exotic) Higgs
doublets or  (exotic) lepton doublets.}.  Unfortunately, the ratio of $b$ and $\tau$
string
scale Yukawa couplings
$1/\sqrt{2}$  is probably  not
consistent with the observed $m_b/m_\tau$ ratio \cite{mbmtau}, and 
the $\tau-\mu$ unification is clearly in disagreement with experiment since
it would lead to approximately equal $\tau$ and $\mu$ masses.
The $t-b$ unification may be acceptable, but only for a sufficiently
large $\tan \beta$, where  $\tan \beta$ is the ratio of VEV's of the
neutral components of the $\bar{h}_c$ and $h_c$ scalars \cite{mbmtau}. (The
MSSM mass relations may be modified because the sum of the squares of
the VEVs of the doublets related to the fermion masses $(h_c, \bar{h}_c)$
may be reduced due to the presence of additional Higgs doublets.) The superpotential also does
 not have Yukawa textures in the
quark sector, so that the first and second quark families (as well as
the first lepton) remain massless.

\item {\bf Running of Gauge 
Couplings:}

From the massless particle content listed in Table IIIb, the $\beta$
functions for the running of the gauge couplings can easily be computed.
As the number of additional $SU(3)_C$ exotic fields is minimal (one
vectorlike pair), the running of $g_3$ is closer to that of the MSSM than
the other gauge couplings.  Therefore, our strategy is to take the value
of $g_3$ at the electroweak scale as an input (we choose $\alpha_s=0.12$ at
$M_Z$), and run the couplings to
the string scale to determine the value of $g=0.80$. (For
our purpose, it is adequate to  consider RGE's at one-loop level,
ignoring SUSY threshold effects.) The other gauge
couplings are then run back to the electroweak scale with this value
of $g$ as an input,
taking into account the Ka\v c-Moody level for the $U(1)$ gauge factors
($k_Y=11/3, k'_Y
\simeq 16.67 
$) and for the hidden sector non-Abelian
groups ($k=2$).

In this case, it would appear from the massless particle content that the
running of the gauge couplings for $SU(2)_L$ and $U(1)_Y$ are very
different from the MSSM case.  However, we find that the low energy values
yield a prediction for $\sin^2\theta_W\sim 0.16$.
While this is lower than the
experimental value ($\sim 0.23$), the disagreement is less than might have
been expected given the
large amount of exotic matter and the value $k_Y=11/3$ (to be compared with the
MSSM value $5/3$). Similarly, for the
$SU(2)$ gauge coupling, we find $g_2=0.48$
surprisingly close to the experimental value 
$\sim0.65$.  The variation of the gauge couplings with the scale is
presented in Figure 2, and the $\beta$ functions are  listed in Table V.
\begin{figure}
\centerline{
\hbox{ 
\epsfxsize=2.8truein
\epsfbox[70 32 545 740]{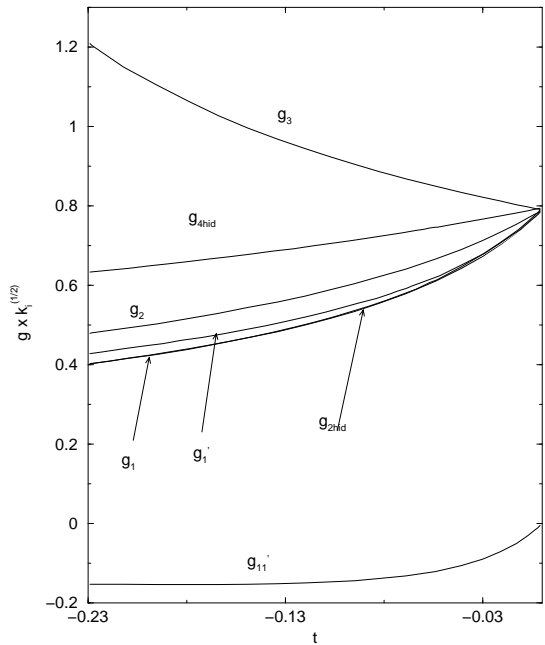}
}
}
\caption{Running of the gauge couplings $\times \sqrt{k}$ 
for the $P_1^{'}P_{2}^{'}P_{3}^{'}$ flat direction, with $t=1/16\pi^2
\ln(\mu/M_{String})$, with $M_{string} =5\times 10^{17}$ GeV,
$g_{|M_{string}}=0.80$. The couplings include the factor $\sqrt{k}$, where $k$
corresponds to the Ka\v c-Moody level (see the caption of Table V for the 
values of $k$).}

%
%
\end{figure} 

In addition, the hidden sector gauge groups are not asymptotically free, and hence
there is no possibility for gaugino condensation or other strong coupling
dynamics
to break supersymmetry in the hidden sector.

\end{itemize}

\section{$P_{2}P_{3}|_{F}$ flat direction}
\label{IV}
The fields involved in this flat direction are $P_{2}P_{3}|_{F}= \{\varphi_{2},
\varphi_{4}, \varphi_{10}, \varphi_{12}, \varphi_{27}, \varphi_{30} \}$
(see Table~II). The
VEV's of these fields are completely fixed due to the $F$- flatness
constraints:
\begin{equation}
|\langle \varphi_{27} \rangle|^2 = 2x^2; \ \  
|\langle \varphi_{30} \rangle|^2=2|\langle \varphi_{2} \rangle|^2= 2|\langle
\varphi_{4} \rangle|^2 =2|\langle \varphi_{10} \rangle|^2= 2|\langle
\varphi_{12} \rangle|^2= x^2 \, ,
\end{equation}
where $x=0.013M_{Pl}$, and $\varphi_{10}$ and $\varphi_{4}$ have
opposite signs of their VEV's.  (Recall, that without loss of generality,
we  take
all the  VEV's real and, except for $\varphi_{10}$, positive.)


The effective mass terms for this direction are 
\begin{eqnarray}
\label{p23Fmassw}
W_{M}&=& gh_f\bar{h}_b \langle \varphi_{27}\rangle +
{g\over {\sqrt{2}}}(e^c_de_b+e^c_ge_a)\langle
\varphi_{30}\rangle + 
{g\over {\sqrt{2}}}\varphi_{3} \varphi_{8} \langle \varphi_{12}\rangle
+{g\over {\sqrt{2}}}\varphi_{8} \varphi_{11} \langle \varphi_{4}\rangle + 
{g\over {\sqrt{2}}}\varphi_{7} \varphi_{16}
\langle \varphi_{2}\rangle  \nonumber \\
&+&{g\over {\sqrt{2}}}\varphi_{1} \varphi_{15} \langle
\varphi_{10}\rangle 
+ \frac{\alpha^{(1)}_{5}}{M_{Pl}^{2}}\varphi_{15} \varphi_{26}
\langle \varphi_{2} \varphi_{4} \varphi_{30}\rangle
+\frac{\alpha^{(2)}_{5}}{M_{Pl}^{2}}\varphi_{7} \varphi_{17} \langle \varphi_{10}
\varphi_{12} \varphi_{30}\rangle \nonumber\\
&+&{g\over {\sqrt{2}}} F_1 \bar{F}_1 \langle \varphi_{30}\rangle +
{g\over {\sqrt{2}}}F_2 \bar{F}_2 \langle \varphi_{30}\rangle 
. 
\end{eqnarray}   
As in the previous case, the coefficients of the elementary trilinear
terms, 
equal to $g$ or  $g/\sqrt{2}$,  are  displayed explicitly. The massive and
massless states are listed in Table IVa and IVb, respectively.  The fields with
nonzero
VEV's   are all massive in this case (the `flat-point' discussed at the
end of Section IIA) and are not included in Table IVa.

After decoupling the heavy fields, 
the effective trilinear couplings for the observable sector states 
are: 
\begin{eqnarray}
\label{w3ob2}
W_{3} &=& gQ_cu^c_c\bar{h}_c + gQ_cd^c_bh_c +
\frac{\alpha^{(3)}_{5}}{M_{Pl}^{2}}Q_ad^c_dh_g \langle \varphi_{4} \varphi_{30}
\rangle + \frac{\alpha^{(4)}_{5}}{M_{Pl}^{2}}Q_bd^c_dh_g \langle \varphi_{12}
\varphi_{30} \rangle + \frac{\alpha^{(5)}_{5}}{M_{Pl}^{2}}Q_bd^c_ch_b
\langle
\varphi_{12} \varphi_{30} \rangle \, 
\nonumber \\
&+& \frac{\alpha^{(6)}_{5}}{M_{Pl}^{2}}Q_ad^c_ch_b
\langle \varphi_{4} \varphi_{30} \rangle 
 + (\frac{\alpha^{(1)}_{4}}{M_{Pl}} u^{c}_b \langle \varphi_{4}
\rangle 
+\frac{\alpha^{(1')}_{4}}{M_{Pl}} u^c_a \langle \varphi_{12}\rangle) d^c_cd^c_d  
\nonumber\\
&+& {g\over {\sqrt{2}}}e^c_ah_ah_c + {g\over {\sqrt{2}}}e^c_fh_dh_c +
(\frac{\alpha^{(2)}_{4}}{M_{Pl}}e^{c}_e \langle \varphi_{12} \rangle  
+\frac{\alpha^{(2')}_{4}}{M_{Pl}}e^c_h  \langle \varphi_{4} \rangle) h_gh_b
\nonumber \\
 &+& \frac{\alpha^{(7)}_{5}}{M_{Pl}^{2}}e_b^ch_gh_a
\langle \varphi_{10} \varphi_{12} \rangle +
\frac{\alpha^{(8)}_{5}}{M_{Pl}^{2}}e_i^ch_gh_a \langle \varphi_{2} \varphi_{4}
\rangle + {g\over {\sqrt{2}}}\bar{h}_ah_c\varphi_{25} +
g \bar{h}_ch_b\varphi_{20} +
g\bar{h}_dh_b\varphi_{28} 
\nonumber \\
 &+&  g\bar{h}_ch_g\varphi_{21} + g\bar{h}_dh_g\varphi_{29}  
\, .
\end{eqnarray}
from which we can redefine two new fields as: 
\begin{equation}
\begin{array}{ccc}
u_{b}^{c'} &=&\frac{M_{Pl}}{\sqrt{(\alpha^{(1)}_{4} \langle \varphi_{4} \rangle)^{2}
+(\alpha^{(1')}_{4} \langle \varphi_{12} \rangle)^{2}}}
(\frac{\alpha^{(1)}_{4}}{M_{Pl}} 
\langle \varphi_{4} \rangle u^{c}_b
+\frac{\alpha^{(1')}_{4}}{M_{Pl}} 
\langle \varphi_{12}\rangle u^c_a); \\
e_{e}^{c'} &=&\frac{M_{Pl}}{\sqrt{(\alpha^{(2)}_{4} \langle \varphi_{12} \rangle)^{2}
+(\alpha^{(2')}_{4} \langle \varphi_{4} \rangle)^{2}}}
(\frac{\alpha^{(2)}_{4}}{M_{Pl}}
\langle \varphi_{12} \rangle e^{c}_e 
+\frac{\alpha^{(2')}_{4}}{M_{Pl}}
\langle \varphi_{4} \rangle e^c_h).
\end{array}
\end{equation}

In addition, there are effective trilinear couplings involving the singlets 
$\varphi_{i}$ and the hidden sector non-Abelian fields which also have
trilinear couplings to the $\varphi_i$. Some of these terms  play
important roles in radiative symmetry breaking scenarios. We  quote only
these terms and the complete discussion of the trilinear terms in the
hidden sector is deferred for further investigation:

\begin{eqnarray}
W_{3hid} &=&\frac{\alpha^{(3)}_{4}}{M_{Pl}} \varphi_{29} \varphi_{21}
\varphi_{25} \langle \varphi_{27} 
\rangle +\frac{\alpha^{(3')}_{4}}{M_{Pl}} \varphi_{28} \varphi_{20}
\varphi_{25} \langle \varphi_{27} \rangle \nonumber\\
&+& \frac{\alpha^{(5)}_{4}}{M_{Pl}}S_2 S_7 \varphi_{21} \langle \varphi_{27}
\rangle + \frac{\alpha^{(6)}_{4}}{M_{Pl}}S_2 S_6 \varphi_{20} \langle \varphi_{27}
\rangle 
\nonumber \\
 &+& \frac{\alpha^{(1)}_{6}}{M_{Pl}^{3}}S_4 S_7 \varphi_{21} 
\langle \varphi_{27} \varphi_{30} \varphi_{12} \rangle +
\frac{\alpha^{(2)}_{6}}{M_{Pl}^{3}}S_8 S_7 \varphi_{21}
\langle \varphi_{27} \varphi_{30} \varphi_{4} \rangle +
\frac{\alpha^{(3)}_{6}}{M_{Pl}^{3}} S_4 S_6 \varphi_{20}
\langle \varphi_{27} \varphi_{30} \varphi_{12} \rangle 
\nonumber \\
 &+& \frac{\alpha^{(4)}_{6}}{M_{Pl}^{3}} S_8 S_6 \varphi_{20}
\langle \varphi_{27} \varphi_{30} \varphi_{4} \rangle \, .  
\end{eqnarray}

\subsection{General Implications: 
}

This model has many features in common with the  previous case: { (i)}
there
are many massless ordinary  fermions and exotics 
for which we have no apparent mechanism to give masses;
{(ii)} the string scale  Yukawa couplings display (unrealistic)  $t-b$ and
$\tau-\mu$
unification
(at least in the scenario where $U(1)'$ is broken at the electroweak
scale) with the  respective Yukawa couplings $g$ and $g/\sqrt{2}$;
(iii) 
there is no  effective ``canonical''  $\mu$ term in the superpotential;
 however,
there is a possibility of a ``non-canonical'' $\mu$ term; {
(iv)} the
$U(1)'$ charges are not family universal.
On the other hand, there are additional features unique to this flat direction.

\begin{figure}
\centerline{
\hbox{
\epsfxsize=6.5truein
\epsfbox[70 32 545 740]{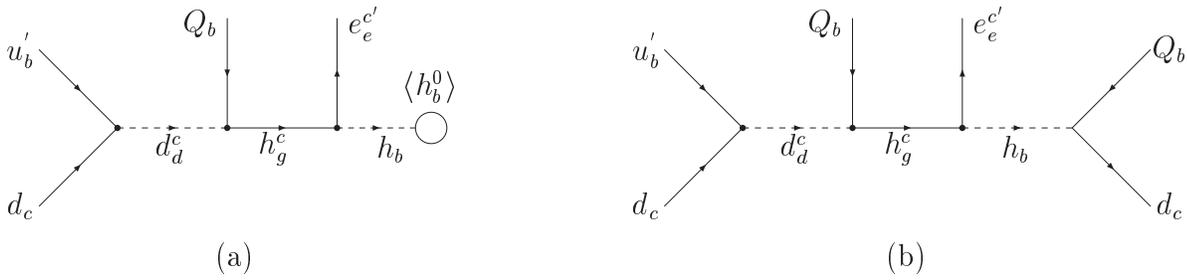}
}
}
\vskip -7.7 truein
\caption{Proton decay diagrams  for  (a) 
$\langle h_b^0 \rangle \ne 0$,
(b) $\langle h_b^0 \rangle = 0$. 
$u_{b}^{'}$, $d_{c}$ are the conjugates of $u_{b}^{c'}$,
$d_{c}^c$.}
\end{figure}

\begin{itemize}

\item {\bf Implications of $L$ and $B$ Violating Couplings:} 

\noindent{\it Proton-Decay.} In this model, there are two  effective
couplings of the type $ud^cd^c$, arising from fourth
order terms in the original superpotential, 
(the  seventh and eighth terms  in eq.
(\ref{w3ob2})). 
The superpotential also has two lepton number violating couplings of the type 
$e^chh$ (the eleventh and twelfth terms in eq. (\ref{w3ob2})), 
since we identify both $h_b$ and $h_g$ as Higgs
doublets due to their couplings to ordinary quarks~\footnote{We could 
instead interpret $h_b$ as a lepton doublet. In that case the $e^{c'}_eh_gh_b$ 
term would conserve  lepton number. However, the $L$-violation would then show up elsewhere,
e.g., the fifth term in (\ref{w3ob2}) would correspond to an $L$-violating
leptoquark interaction.}.  As a result of these
two types of
couplings, there is potential danger for proton decay,
illustrated by the diagrams  
displayed 
in Figure 3~\footnote{For some other flat directions there is,  along
with the baryon violating term of the type $ud^cd^c$, also 
a lepton violating term of the type $Qd^ch$ (where $h$ is a lepton doublet),
e.g.,  the $P_1P_2P_3|_F$ direction has such a term.  In this case
 proton-decay takes place via an effective dimension six operator
and is even harder to suppress.}.   The decay rate
depends on the masses of the fields involved, and on whether $h_b^0$ has a VEV,
 and thus it  depends
on the details of the soft breaking as well as on a particular 
identification of the
particles participating in the process; i.e., some participants can be
identified
with the exotics  and/or the  second and third family fermions.
 In particular, $e_e^{c'}$
 could be exotic, which would suppress the decay. However, this state is 
  massless within our approach, i.e., the model is not 
  sufficiently realistic to make many of the exotic  
  states significantly massive.   

The proton decay rate will be much too fast unless some of the external legs almost 
completely decouple from the first two families.
 Defining $U_a^{(4)}$ as the product
of the unitary matrix elements relating the four external legs in 
Figure 3a to the 
states relevant to proton decay, and $U_b^{(6)}$ as the corresponding product of six 
matrix elements for Figure 3b, one requires
\begin{eqnarray}
\left\langle h_b^0\right\rangle U_a^{(4)} &<& 10^{-24} m_{prop}^3\, ,\\
U_b^{(6)} &<& 10 ^{-15} m_{prop}^5\, ,
\nonumber
\end{eqnarray}
where $\left\langle h_b^0\right\rangle$ and $m_{prop}$
 (a  mass scale for 
the internal propagators, typically set by the electroweak/SUSY scale) are
in TeV and we have assumed a lifetime of $\tau >10^{34}
{\rm yr}$.

\begin{figure}
\centerline{
\hbox{
\epsfxsize=6.0truein
\epsfbox[70 32 545 740]{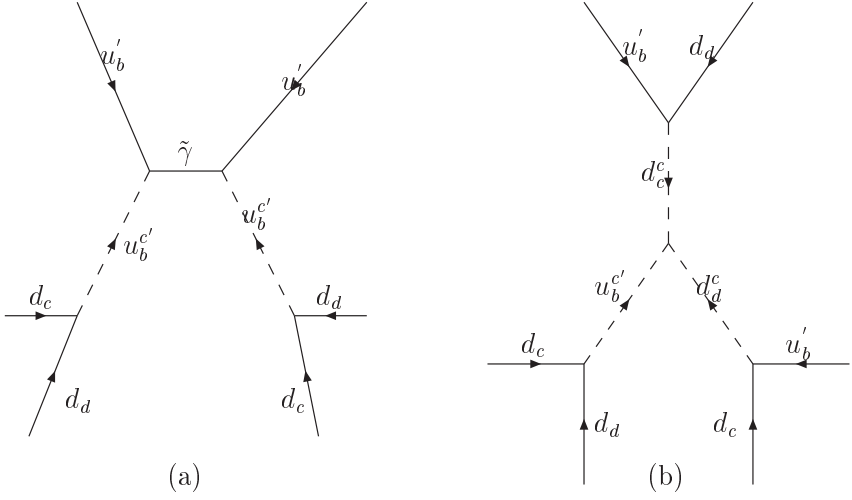}
}
}
\vskip -5.7 truein
\caption{Diagrams for the $\Delta B=2$ processes $N\rightarrow \bar{N}$
(neutron oscillation) or $\Delta B=2$ nuclear decay.  The tri-scalar
vertex in (b) corresponds to  a soft supersymmetry 
breaking ``$A$-term''.
 $u_{b}^{'}$, $d_{c}$ and $d_{b}$ are the congugate of $u_{b}^{c'}$,
$d_{c}^c$ and $d_{b}^c$.
}
\end{figure}

\noindent{\it $N-\bar N$ Oscillations.} Even if proton decay is
somehow suppressed, the coupling of the type $ud^cd^c$
also implies that there are  $N-\bar N$ oscillations 
 via a diagram  that
involves   gaugino exchange (see Figure 4a).
 Another process  that can
contribute involves exchanges of three virtual squarks (Figure 4b), but is
more model dependent since it depends on the trilinear soft supersymmetry
breaking   term associated with $ud^cd^c$. (For a  recent review
of $N-\bar N$ oscillations see~\cite{Mohap}.) Since all 
 the effective  trilinear terms  in the superpotential are fixed for this
flat direction,  and this
coupling appears at the fourth order and is {\it not} suppressed,  the
only suppression is due  to the identification of the fields in the
coupling with exotics and/or second and
 third family fermions. (For some other flat directions, e.g.,
$P'_1P_2P_3|_F$, the trilinear coupling depends on one free VEV, allowing
for additional suppressions if that VEV is small.)

From the experimental limits on oscillations 
and $\Delta B=2$ nuclear decays \cite{Mohap}, 
one finds
\begin{equation}
U_c^{(6)} \simlt 10^{-11} m_{prop}^5,
\end{equation}
where $U_c^{(6)}$ is the product of the unitary matrix elements 
for the six external fermions 
onto the neutron states and $m_{prop}$ is in TeV.  Again one would need
one or more of the states to almost decouple from the first family.

\noindent{\it $R$-parity Violating Processes.} The lepton- and baryon-
 violating couplings imply that there is no stable  lightest supersymmetric
 particle (LSP). E.g., if one  assumes
 that the LSP is a gaugino, the baryon- violating coupling 
  allows for a decay of a gaugino  into three quarks  via the exchange of
a virtual squark, as shown in Figure 1b; 
  and the lepton- violating
  coupling into  three leptons via a virtual slepton, by a diagram
analogous to Figure 1a.

\item {\bf Textures:} 

The Yukawa couplings for the ordinary quarks display a possible texture
for the down-type quarks, if the Higgs doublets $h_c$,
$h_b$, and $h_{g}$ all acquire VEV's. In this
case, the mass matrix for the down-type quarks is \footnote{It is possible
for $\langle h_g \rangle$ and $\langle h_b \rangle$ to accquire phases by
a spontaneous breaking of CP for some symmetry breaking patterns
\cite{cceelw2}. However, these phases can be absorbed into the definition
of the mass eigenstate quarks, so that $U_{CKM}$ is real for this model. } 
\begin{equation}
\label{massmatrix}
M=\left ( 
\begin{array}{ccc}
\langle h_g \rangle \frac{\alpha_5^{(3)}x^2}{\sqrt{2} M_{Pl}^2} &
\langle h_b \rangle \frac{\alpha_5^{(6)}x^2}{\sqrt{2} M_{Pl}^2} & 
0\\
\langle h_g \rangle \frac{\alpha_5^{(4)}x^2}{\sqrt{2} M_{Pl}^2} & 
\langle h_b \rangle \frac{\alpha_5^{(5)}x^2}{\sqrt{2} M_{Pl}^2} &
0 \\
0 & 0 & g\langle h_c \rangle 
\end{array} \right ),
\end{equation}
and the mass terms in the Lagrangian are
\begin{equation} 
-L_{mass}=(Q_{a}, Q_{b}, Q_{c}) M \left (
\begin{array}{c}
d_{d}^{c} \\
d_c^{c} \\
d_b^{c}
\end{array} \right ). 
\label{texture}\end{equation}
The state $d^c_{a}$ is the partner of the exotic ${\cal D}_a$; both
remain massless in this model.
It may be possible for all four Higgs doublets
$\bar{h}_c$, $h_{c}$, $h_{b}$ and $h_{g}$ to acquire non-zero VEV's. Then
the mass matrix
for the down-type quarks has the most general form given in 
(\ref{massmatrix}), which by inspection demonstrates that there is no
mixing between the third family and the first two families.  In general,
there are three massive states: one heavy (bottom) quark, and two lighter
quarks due to suppressions from the higher order terms. One of them may
be 
massless if the string coefficients
$\alpha_{5}^{(3)}=\alpha_{5}^{(4)}$ and 
$\alpha_{5}^{(5)}=\alpha_{5}^{(6)}$, which also results in 
nearly maximal mixing in the case in which $\langle
h_b \rangle \sim \langle h_g \rangle$.  If either
$h_b$ or $h_g$ does not acquire a VEV, there will be one massless and two
massive eigenstates (once again, with a hierarchy of masses).  
Clearly, if both $h_b$ and $h_g$ do not acquire VEV's, the down quark
states of the first two families will remain massless. 

It is convenient to define the dimensionless ratios (for the case $\langle
h_c\rangle\ne 0$):
\beqn
\kappa_{3,4} &  = &
\frac{\langle h_g \rangle}{\langle h_c \rangle}
\frac{\alpha_5^{(3,4)}x^2}{\sqrt{2} g M_{Pl}^2} 
\sim 0.1 \frac{\langle h_g \rangle}{\langle h_c \rangle}\,, \nonumber \\
\kappa_{5,6} &  = &
\frac{\langle h_b \rangle}{\langle h_c \rangle}
\frac{\alpha_5^{(5,6)}x^2}{\sqrt{2} g M_{Pl}^2}
\sim 0.1 \frac{\langle h_b \rangle}{\langle h_c \rangle},  \eeqn
where the numerical values use the estimates of $\alpha_5$
from Section II. Then, neglecting the running of the Yukawas
down to low energies, the quark masses are
\beqn
(m_u,m_c,m_t) & = & (0,0,1) \times g \langle \bar{h}_c \rangle\,, \nonumber \\
(m_d,m_s,m_b) & = & (\delta_1,\delta_2,1)
\times g \langle h_c \rangle,  \eeqn
in which $\delta_{1,2}$ are the mass eigenvalues
\beqn
\delta_{1,2}=\left [\frac{\sum_{i=3}^{6}\kappa_i^2}{2} \mp
\frac{1}{2}((\kappa_6^2+\kappa_3^2-\kappa_4^2-\kappa_5^2)^2+4(\kappa_6\kappa_5 
+\kappa_3\kappa_4)^2)^{1/2} \right ]^{1/2}.
\eeqn  
Although this is not fully realistic, it illustrates the
possibility of a realistic $m_s/m_b$  ratio due to  the contribution of
the fifth order terms. For $\kappa_3 = \kappa_4$ and $\kappa_5 =
\kappa_6$, one has $m_d=0$, as expected. However, a small $m_d / m_s$ can
emerge if $\kappa_3 \neq \kappa_4$ or $\kappa_5 \neq \kappa_6$. One does
not expect the $\alpha$'s of the same order to be equal in general,
leading to the possibility of a small $m_d /m_s$ even though they both
are from dimension 5 operators.  
The expected ratios of the doublet VEV's will
be discussed in \cite{cceelw2}. It is straightforward to determine the
CKM quark mixing matrix corresponding to this texture.
One finds
\be
U_{CKM} = \left(  \begin{array}{ccc}
 \cos\theta_c & \sin\theta_c &  0 \\
 -\sin\theta_c & \cos\theta_c  &  0 \\
 0 & 0  & 1
\end{array} \right),  \ee
in which the mixing angle $\theta_c$ is given by
\be
\tan(2 \theta_c)=\frac{2(\kappa_3\kappa_4+\kappa_5\kappa_6)}{\kappa_3^2 
+\kappa_6^2 -\kappa_4^2-\kappa_5^2}.
\ee

For $\kappa_3 = \kappa_4$ and $\kappa_5 = \kappa_6$, one obtains maximal
mixing and $\theta_{c}=\pi/4$. Relaxing the equality of the $\alpha_{5}$'s
one obtains a more realistic value for $\theta_{c}$, as well as $m_d/m_s
\neq 0$. However, the relation $\theta_{c} \sim (m_d/m_s)^{1/2}$ ($\sim
0.2$) does not hold except for special values of the parameters. 

As for the texture in the lepton sector,
if $h_g$ acquires a non-zero VEV  then the  terms  $e^c_bh_gh_a$ and
$e^c_ih_gh_a$ in (\ref{w3ob2})  can  (slightly) break the  (unrealistic)
$\tau-\mu$ degeneracy. However, 
 the electron remains massless in this scenario.

\item {\bf Running of the Gauge Couplings:}

The mass spectrum for the quark sector for this model is the same as in  
the previous case; hence, we adopt the same strategy for the running of
the gauge couplings. The $\beta$ functions, as listed in Table V, are
similar to those of the $P_1'P_2'P_3'$ direction. Again we 
 choose $\alpha_s=0.12$ at
$M_Z$ and run the couplings to
the string scale to determine the same value of $g=0.80$. (The number of
massless colored fields is the same as in the previous example.)
We find the values
$g_1=0.40$ (which includes $k_Y=11/3$ factor) and $g_2=0.46$ ($k_2=1$) 
at the electroweak scale, yielding
$\sin^2\theta_W=0.17$.     

\end{itemize}
\subsection{Possible $U(1)'$  Symmetry Breaking Scenarios}

In this model, $\varphi_{25}$ remains massless at the string scale (see
Table IVb). 
 This field, which has $U(1)'$ charge opposite in sign to all the other
singlet fields (with nonzero $U(1)'$ charge), is required for ensuring the
$U(1)'$ 
$D$- flat direction, and thus is necessary for  the
intermediate scale symmetry breaking scenario, as discussed in
\cite{cceel1}. The $U(1)'$  $D$- flat direction involves
$\varphi_{25}$
and one of the singlet fields $\{ \varphi_{18},
\varphi_{19},\varphi_{20},\varphi_{21}, 
\varphi_{22} \}$. The couplings of these singlet fields seem to 
indicate that an intermediate scale flat direction involving
$\varphi_{20}$ or $\varphi_{21}$ is potentially dangerous, as it would
decouple $\bar{h}_c$, and hence the top
quark coupling.  
However, the $D$- flat directions involving $\varphi_{20}$,
$\varphi_{21}$ are not $F$- flat at the renormalizable level. In
particular, they are lifted by the couplings
$\varphi_{25}\varphi_{20}\varphi_{28}$ and  
$\varphi_{25}\varphi_{21}\varphi_{29}$, respectively, and thus the
symmetry breaking scale would take place at the electroweak scale.

If one can ensure that the supersymmetry breaking 
mass-squares of $\varphi_{20}$ and $\varphi_{21}$  stay positive
 at low energies, while $m_{\varphi_{25}}^{2}+m_{\varphi_{18/19/22}}^{2}$ 
 is negative, then  we can have intermediate scale symmetry breaking along
the $D$-flat direction(s) $\varphi_{25}+\varphi_{18/19/22}$.

Using gauge invariance arguments and string selection rules, we find that
there are no higher dimensional non-renormalizable operators (NRO's) in
the superpotential involving self-couplings of the singlets in
the intermediate scale $D$-flat direction, 
which in principle could stabilize the flat direction and 
determine the intermediate scale of the symmetry
breaking. Hence, in this model the symmetry breaking is purely radiative,
with the  scale $\mu_{RAD}$ very close to the scale at which the
sum of the mass squares
$m_{\varphi_{25}}^{2}+m_{\varphi_{18/19/22}}^{2}$ crosses
zero~\cite{cceel1}. 

After the intermediate scale symmetry breaking, the electroweak symmetry
breaking has a few  novel features, different from  the previous case:

\noindent(i)$~$ In this case,  $h_c$ is heavy and decouples,
and so the bottom quark Yukawa is absent in the theory.
The first and second families can have electroweak scale masses if $h_b$
and $h_g$ acquire VEV's.
In addition, the decoupling of $h_c$ makes $\mu$ and $\tau$
massless at the electroweak 
scale, since couplings $h_ah_ce_a^c$ and $h_dh_ce_f^c$ all disappear.     

\noindent(ii)$~$The running of the gauge couplings has  to be modified in order 
 to take into
account the decoupling of the heavy states. However, the
complete determination of the fields with intermediate scale
masses requires a detailed knowledge of the relevant NRO's, and is beyond the
scope
of the paper.

\noindent (iii)$~$In principle, it  is also possible to give the total
singlets (with $Q_{Y}=Q^{'}=0$)
intermediate scale VEV's; since  they do not have any $D$- term, the
symmetry breaking scenario  is similar to the  case with the charged SM
singlets in a $D$-flat direction. In this model, the only total singlets
which have Yukawa couplings (and hence their mass-squares can be driven
negative radiatively) are $\varphi_{28}$ and $\varphi_{29}$.  However,
these fields couple to $\varphi_{25}$ at the effective trilinear
level, which give rise to effective $F$- terms that push the VEV's down to
the electroweak scale.   
Therefore, $\varphi_{28}$ and $\varphi_{29}$ cannot be involved in the
intermediate scale flat direction at the same time as $\varphi_{25}$.

These possibilities are discussed in detail  for  specific Ans\"atze for the
soft supersymmetry breaking  mass 
terms  in \cite{cceelw2}, while the implications for
generating  the effective $\mu$ term as well as  ordinary and exotic
fermion masses via higher dimensional operators \cite{cceel1} is deferred
to a future study.


 \section{Conclusions}\label{V}

In this paper we have given a thorough
 investigation of the  physics implications for the  observable sector of  
the CHL5
model from the ``top-down'' approach. For this model a complete
classification of the $D$- and $F$- 
flat directions  due to non-zero VEV's of the non-Abelian singlets 
was found~\cite{cceel3}; these directions were shown to be flat to all
orders in string
perturbation.  Along all these flat directions there is at least one
additional $U(1)'$  factor (along with the SM gauge group).
We chose two particular flat directions
as representative examples which exhibit the type of physics phenomena
characteristic for the  quasi-realistic string vacua based on free-fermionic
constructions. 

For each of the representative directions we found a complete massless spectrum
and the trilinear terms in the observable sector of the 
 superpotential. (We also determined the terms of the observable
sector fields to the hidden ones that may play a  role in the RGE
analysis.) Gauge invariant 
 bilinear and trilinear terms were first
determined to all orders in  fields with non-zero VEV's  along a
particular flat direction.
The subsequent string calculation  for a particular gauge
 invariant term then determines whether this term  is indeed there. 
 Notably, the world-sheet constraints of
 string theory disallow  a large number of gauge invariant terms;
  thus, there are in
 general fewer terms than expected.
 These ``string selection-rules'' have in general important physics 
implications:
 
\begin{itemize}

\item {\bf Massless spectrum.}
Along with the MSSM particle content there are  a
large number of additional massless (at the string scale) exotics. 
In the case of electroweak scale $U(1)'$ breaking, the exotic
fermions are light compared with the electroweak scale, which is clearly
excluded by experiment. This feature seems to pose a serious problem in
the search for a realistic string vacuum.

Surprisingly, the type of massless particle content
that survives for the two representative flat directions,
combined with the higher Ka\v c-Moody level for $U(1)_Y$,
still allows for a gauge coupling unification with a prediction
for $\sin\theta_W \sim 0.16$ that, while not consistent with experiment,
is not too far away from the experimental value.

  \item{\bf Trilinear Couplings.}
The string  models based on a free fermionic construction
possess
the feature that  nonzero trilinear terms at the third order are 
of the 
order of the gauge coupling, and thus
large. For the particular model discussed they are  equal to $g$ or
$g/\sqrt{2}$. (Such large trilinear
terms facilitate the radiative symmetry breaking scenarios of the SM and 
$U(1)'$  gauge structure.)  Importantly, 
the trilinear terms  surviving at the fourth order  turn out to  have
effective couplings that are comparable to those at the third
order~\cite{CEW}. Only at the
fifth and higher orders, the suppression of these coupling becomes
significant, i.e., of order 1/10 and smaller.   

Again, the number of allowed trilinear terms in the
superpotential is significantly smaller than allowed by gauge invariance,
which has a number of  implications:

\noindent{\it Fermion Textures.}
The CHL5 model possesses 
a distinct feature that  in the quark sector only one up-type and one
down-type quark have Yukawa couplings 
at the third order, while in the lepton sector generically two $e$-type
couplings have Yukawa couplings at the third order,  
so that at the string scale there is  
$t-b$ and $\tau-\mu$ universality with the respective couplings $g$ and
$g/\sqrt{2}$.  While $t-b$ universality is
consistent with experiment for sufficiently large $\tan \beta$,
 the ratio of $b$- and $\tau$- Yukawa couplings is somewhat large,
and $\tau- \mu$  universality is clearly not physical.
For certain  flat directions 
(e.g., the  first example) there is
no further  texture in the fermion mass matrix. On the other hand, in
other directions (e.g., the second example)
the texture is induced at a higher order -- however,  
only in the down quark- sector, and only certain entries arising at a
specific order (e.g, for the second example, at the fifth
order).
These features are modified in the intermediate scale case, in which
the $h_c$ acquires an intermediate scale mass (and no VEV), and
most of the fermion masses would have to be generated by higher
dimensional operators.

\noindent{\it $\mu$ parameter.} In principle, there is a possibility of
having a  standard effective $\mu$-term at the electroweak scale, due to the
trilinear couplings of the  two standard Higgs doublets
(which respectively couple to the $t$- and $b$- quarks) to the SM
singlet(s) which   acquire  non-zero VEV's at the
electroweak scale (and, if they  are charged under the additional $U(1)'$,
give mass to the $Z'$). It turns out that
there is no such canonical  $\mu$-term in these examples. However,
there are ``non-canonical'' $\mu$-terms that involve a standard Higgs
doublet with a 
coupling  to  the $t$-quark and another Higgs field (that may couple to
the $s$- or $d$-quark at high orders). This non-canonical $\mu$-parameter
has interesting implications for the neutralino and chargino sector
mass patterns.

\noindent{\it  Baryon and lepton-number violating couplings.} In
general  such couplings are present; they may induce
proton decay, $N-{\bar N}$ oscillations and/or leptoquark couplings and break
$R$-parity, so that there is no stable LSP. For  specific
directions such  couplings are  absent or could be suppressed within 
a  specific SM symmetry breaking scenario.

\item{\bf The $U(1)'$ symmetry breaking pattern} can be either at an
intermediate or at the electroweak scale. It depends on the 
 $U(1)'$ charges and the type of trilinear couplings of the massless  SM
singlets.
E.g., for the first representative direction
all the $U(1)'$ charges of  the massless SM singlets have the same sign; 
thus,  there is  no $D$-flat  $U(1)'$ direction, 
 and the breaking
necessarily takes place at the electroweak scale. 
The particular values of the  $U(1)'$ charges for the light
particle spectrum (which are family non-universal), imply new experimental 
constraints on this $Z'$.
On the other hand, in the
second example there is an additional massless SM singlet with the
opposite sign
of $U(1)'$ charge from the other singlets, and  $F$-flat
$U(1)'$-directions for the trilinear couplings, and thus the symmetry
breaking can take place at an intermediate scale.
It turns out to be  purely radiative in origin,  because of the absence of
the relevant nonrenormalizable terms in the superpotential.

\end{itemize}

Further specific properties of the symmetry breaking patterns,
and 
the corresponding mass spectrum  at the
electroweak
scale, depend on the soft supersymmetry
breaking mass terms  introduced by hand at the string scale.
Specific examples
with a realistic  electroweak SM  and $U(1)'$ breaking pattern,
will be investigated in  \cite{cceelw2}.

While the trilinear couplings in the hidden sector were not explored  to
all orders in the VEV's of the fields in a particular flat direction, the
terms at the third and fourth order are known~\cite{chl}. 
(Terms at higher orders are suppressed.) 
Further investigation of the implications of the hidden sector is
underway.

 The work presented in this paper opens a number of further avenues for
investigation.
Specifically, the study of non-renormalizable terms in the superpotential
is needed. In this case the investigation is complicated by the fact that
along with the direct determination of the non-renormalizable terms in
string theory   there are additional induced terms 
due to the decoupling effects of the fields that became heavy
after vacuum restabilization.  (For detailed investigation
of  these decoupling effects see \cite{ceew}.) 
Further study
of the corrections to the K\" ahler potential after vacuum
restabilization is also needed.
These effects may have further  implications 
for the mass spectrum  in the  intermediate scale $U(1)'$ breaking
scenarios. In particular, a number of exotic fields could acquire mass at
a scale  larger than the electroweak one. In
addition, additional entries  in the fermion textures can
appear and effective $\mu$ term can be generated~\cite{cceel1}.

The techniques for systematic classification of   $D$- and
$F$- flat
directions \cite{cceel2,cceel3} for perturbative heterotic string vacua
with anomalous $U(1)'$
and the subsequent determination of
the mass spectrum and coupling of the restabilized vacua, as investigated
in this paper, is 
general, and can be applied immediately to the study of
other quasi-realistic models, which is also underway. 
These techniques   may also be
applied to  the study of 
{\it non-perturbative} heterotic string vacua  with anomalous
$U(1)'$~\cite{lykken}.

The models discussed in this paper are not fully realistic, and
contain such features as very light or massless exotic fermions,
charginos, and neutralinos, an unwanted $\tau-\mu$ universality and
undesirable ratio of the $b$- and $\tau$ string scale Yukawa couplings,
unrealistic fermion textures, and possible proton decay, etc.
However, they also contain at least the gauge structure and particle
content of the MSSM, and illustrate a
number of features of this class of string models that are likely 
to be present in many other string models, including nonperturbative ones.
These include additional
$Z'$ bosons, which may have family-nonuniversal couplings and
which may have masses
either at the electroweak scale or an intermediate scale; exotic fermions
and their scalar partners; approximate gauge unification; 
the possibility of effective  non-standard $\mu$ terms;
an extended neutralino/chargino spectrum;
 the possibility of baryon and/or lepton number violation;
and $R$-parity violation, leading to the absence of a stable LSP,
leptoquark couplings, 
and non-standard chargino/neutralino decays.

\acknowledgments
We would especially like to thank J. Lykken for making available to us the 
program that generates the massless spectrum of free fermionic
string vacua as well as the original program that calculates the
superpotential couplings.  We would also like to thank him for numerous
discussions and guidance that enabled us to determine fully the relevant
superpotential couplings. M.C., L.E. and J. W. would also like to thank
Fermilab, where part of the work was done, for hospitality.
L.E. also thanks Gordon Kane for helpful discussions.  This work was supported in
part by U.S. Department of Energy Grant No. 
DOE-EY-76-02-3071. 
\newpage

\def\B#1#2#3{\/ {\bf B#1} (19#2) #3}
\def\NPB#1#2#3{{\it Nucl.\ Phys.}\/ {\bf B#1} (19#2) #3}
\def\PLB#1#2#3{{\it Phys.\ Lett.}\/ {\bf B#1} (19#2) #3}
\def\PRD#1#2#3{{\it Phys.\ Rev.}\/ {\bf D#1} (19#2) #3}
\def\PRL#1#2#3{{\it Phys.\ Rev.\ Lett.}\/ {\bf #1} (19#2) #3}
\def\PRT#1#2#3{{\it Phys.\ Rep.}\/ {\bf#1} (19#2) #3}
\def\MODA#1#2#3{{\it Mod.\ Phys.\ Lett.}\/ {\bf A#1} (19#2) #3}
\def\IJMP#1#2#3{{\it Int.\ J.\ Mod.\ Phys.}\/ {\bf A#1} (19#2) #3}
\def\nuvc#1#2#3{{\it Nuovo Cimento}\/ {\bf #1A} (#2) #3}
\def\RPP#1#2#3{{\it Rept.\ Prog.\ Phys.}\/ {\bf #1} (19#2) #3}
\def\etal{{\it et al\/}}

\bibliographystyle{unsrt}

\begin{references}

\bibitem{CY} P. Candelas, G.T. Horowitz, A.
Strominger and E. Witten, {\it  Nucl. Phys.} {\bf B258} (1985)46.
\bibitem{lat}W. Lerche, D. Lust and A.N. Schellekens,  
{\it Phys. Lett.} {\bf 181B}  (1986) 71 and {\it Nucl. Phys.} {\bf B287}
(1987) 477.
    \bibitem{ABKW}{I. Antoniadis, C. Bachas, C. Kounnas, and P. Windey, 
\PLB{171}{86}{51};\\
I. Antoniadis, C. Bachas, and C. Kounnas, \NPB{289}{87}{87};
I. Antoniadis and C. Bachas, \NPB{298}{88}{586}.}
%
\bibitem{KLST}{H. Kawai, D. Lewellen, and S.-H. Tye, \NPB{288}{87}{1};
H. Kawai, D. Lewellen, J.A. Schwartz, and S.-H. Tye, \NPB{299}{88}{431}.}
\bibitem{DHVW}{L.~Dixon, J.A.~Harvey, C.~Vafa and E.~Witten,
\NPB{261}{85}{678} and \B{274}{86}{285};
 L.~Ib\'a\~nez, H.P.~Nilles and F.~Quevedo,
\PLB{187}{87}{25} and \NPB{192}{87}{332}.
%
\bibitem{orbifolds} 
L.~Ib\'a\~nez, J.E.~Kim, H.P.~Nilles and F.~Quevedo, \PLB{191}{87}{282};
J.A.~Casas and C.~Mu\~noz, \PLB{209}{88}{214} and \B{214}{88}{157}; 
J.A.~Casas, E.~Katehou and C.~Mu\~noz, \NPB{317}{89}{171}; 
A.~Font, L.~Ib\'a\~nez, H.P.~Nilles and F.~Quevedo,
\PLB{210}{88}{101};
A.~Chamseddine and M.~Quir\'os, \PLB{212}{88}{343},
\NPB{316}{89}{101};
A.~Font, L.~Ib\'a\~nez, F.~Quevedo and A.~Sierra,
\NPB{331}{90}{421}.
%
\bibitem{calabiyau} B.~Greene, K.~Kirlin, P.~Miron and G.G.~Ross,
\NPB{278}{86}{667} and \B{292}{87}{606}


\bibitem{NAHE} 
I. Antoniadis, J. Ellis, J. Hagelin, and D.V. Nanopoulos,
 \PLB{231}{1989}{65};  I. Antoniadis, G.K.
Leontaris and  J. Rizos, {\it Phys. Lett.} {\bf B245} (1990) 161;
A.E. Faraggi, \NPB{387}{92}{239};
  [hep-th/9708112]; A.E. Faraggi and D.V. Nanopoulos, \PRD{48}{93}{3288}.

\bibitem{FNY1}{A. Faraggi, D.V. Nanopoulos, and K. Yuan, \NPB{335}{90}{347};
A. Faraggi, \PRD{46}{92}{3204}.} 

\bibitem{AF1}{A. Faraggi, \PLB{278}{92}{131}, 
\NPB{403}{93}{101} and \PLB{339}{94}{223}.} 

\bibitem{chl}{S. Chaudhuri,  G. Hockney and J. Lykken, \NPB{469}{96}{357} 
and Erratum-ibid, in preparation.}
\bibitem{Kap}{V. Kaplunovsky, \NPB{307}{88}{145} and  Erratum-ibid. {\bf
B382} (1992)436.}
\bibitem{dienes} K.R. Dienes, Phys. Rept. {\bf 287}, 447 (1997).
\bibitem{CFTO}{L.~Dixon, E.~Martinec, D.~Friedan and S.~Shenker,
\NPB{282}{87}{13}.}
%
\bibitem{C}{L.~Dixon and M.~Cveti\v c, unpublished; M.~Cveti\v c, in 
Proceedings of {\it Superstrings, Cosmology, and
Composite Structures}, College Park, Maryland, March 1987, S.J.~Gates and 
R.~Mohapatra, eds. (World Scientific, Singapore, 1987) and
\PRL{59}{87}{2829}.}
\bibitem{worldsheet}{D. Bailin, D. Dunbar, and A. Love,
\PLB{219}{89}{76};
S. Kalara, J. L\'opez, and D.V. Nanopoulos,  
\PLB{245}{90}{421}, \NPB{353}{91}{650};
A. Faraggi, \NPB{487}{97}{55}.}

\bibitem{cceel2}{G. Cleaver, M. Cveti\v c, J.R. Espinosa,
L. Everett, and P. Langacker, \NPB{525}{98} 3.} 
\bibitem{cceel3} {G. Cleaver, M. Cveti\v c, J.R. Espinosa,
L. Everett, and P. Langacker, [hep-th/9805133], submitted to {\it
Nucl.\ Phys.\ }{\bf B}.}

\bibitem{DSW}{M. Dine, N. Seiberg, and E. Witten, \NPB{289}{87} 585.}

\bibitem{ADS}{J. Atick, L. Dixon and A. Sen, \NPB{292}{87}{109};
M. Dine, I. Ichinose, and N. Seiberg, \NPB{293}{87}{253};
M. Dine and C. Lee, \NPB{336}{90}{317}.}

\bibitem{AS}{J. Atick and A. Sen, \NPB{296}{88}{157}.}

\bibitem{DK}{L. Dixon and V. Kaplunovsky, unpublished.}

\bibitem{genanomp}{
J.A. Casas and C. Mu\~noz, \PLB{216}{89}{37};
M. Yamaguchi, H, Yamamoto, and T. Onogi,
 \NPB{327}{89}{704};
J.A. Casas, J.M. Moreno, C. Mu\~noz, and M. Quir\'os,
\NPB{328}{89}{272};
J. L\'opez and D.V. Nanopoulos, \PLB{245}{90}{111};
C.J.H. Lee, Ph.D. thesis, UMI-92-24830;
A. Faraggi, \NPB{387}{92}{239}, \PLB{302}{93}{202}, \NPB{407}{93}{57}; 
H. Nakano, KUNS-1257 [hep-th/9404033],
T.~Kobayashi and H.~Nakano, \NPB{496}{97}{103},
G.~Cleaver and A.E.~Faraggi, UFIFT-HEP-97-28, UPR-0773-T, hep-ph/9711339
.}
\bibitem{genanoms}{
Y. Kawamura and T. Kobayashi, \PLB{375}{96}{141};
A. Faraggi and E. Halyo, \IJMP{11}{96}{2357};
G. Dvali and A. Pomarol, \PRL{77}{96}{3728} and [hep-ph/9708364];
P. Bin\'etruy and E. Dudas, \PLB{389}{96}{503};
R.N. Mohapatra and A. Riotto, \PRD{55}{97}{1138},
\PRD{55}{97}{4262}; 
Z. Berezhiani and Z. Tavartkiladze, \PLB{396}{97}{150};
K. Choi, E. Chun, and H. Kim, \PLB{394}{97}{89}; 
P. Bin\' etruy, N. Irges, S. Lavignac, and P. Ramond, \PLB{403}{97}{38};
R.J. Zhang, \PLB{402}{97}{101};
A. Nelson and D. Wright, \PRD{56}{97}{1598};
M.M. Robinson and J. Ziabicki, EFI-97-22 [hep-ph/9705418];
 N.K. Sharma, P. Saxena, S. Smith, A.K. Nagawat, and R.S. Sahu,
\PRD{56}{97}{4152};
N. Irges and S. Lavignac, UFIFT-HEP-97-34 [hep-ph/9712239];
A. Faraggi and J. Pati, UFIFT-HEP-97-29 [hep-th/9712516];
A. Faraggi, UFIFT-HEP-98-5 [hep-ph/9801409];
N. Irges, S. Lavignac, and P. Ramond, UFIFT-HEP-98-06 [hep-ph/9802334];
A. Pomarol, PRINT-98-021.}

%

\bibitem{ceew}{M. Cveti\v c,  L. Everett, and J. Wang, 
[hep-ph/9807321].}
%
\bibitem{witten}{E. Witten,  {\it Nucl. Phys.} {\bf B268}, 79
(1986);
 M. Dine and  N. Seiberg, {\it Phys. Rev. Lett.} {\bf 57}, 2625 (1986).}
%
\bibitem{cceelw2}{G. Cleaver, M. Cveti\v c, J.-R. Espinosa, L. Everett,
J. Wang and P. Langacker, UPR-814-T, in preparation.}
%

%
%
%
%
\bibitem{cl}{M.~Cveti\v c and P.~Langacker,
\PRD{54}{96}{3570}, \MODA{11}{96}{1247} and [hep-ph/9707451].}
%
\bibitem{SY}{ D. Suematsu and Y. Yamagishi, Int. J. Mod. Phys {\bf A10} 
(1995) 4521.} 
 \bibitem{cdeel}{M.~Cveti\v c, D.A.~Demir, J.R.~Espinosa, L.~Everett and 
P.~Langacker, \PRD{56}{97}{2861}; G.~Cleaver {\it et al.}, Ref.~\cite{cceel1}.
}
%
\bibitem{lw}{P. Langacker and J. Wang,
[hep-ph/9804428], to be published in 
{\it Phys. Rev.} {\bf D}.}
\bibitem{KM}{E. Keith and E. Ma,  {\it Phys. Rev.} {\bf D56} (1997) 7155.}
\bibitem{cceel1}{G. Cleaver, M. Cveti\v c, J.R.
Espinosa, L. Everett, and  P. Langacker, {\it Phys. Rev.} {\bf 
D57}, 2701 (1998).}
%
%
%
%
%
%
%
%
%
\bibitem{flat1} See, e.g., M.A.~Luty and W.~Taylor IV, {\it Phys. Rev.}
{\bf D53}, 
(1996) 3399 and references therein.
%
\bibitem{flat2} T.~Gherghetta, C.~Kolda and S.~Martin, \NPB{468}{96}{37}.
%
\bibitem{flat3} P.~Bin\'etruy, N.~Irges, S.~Lavignac and P.~Ramond,
\PLB{403}{97}{38}.
%
%
\bibitem{CEW}{M. Cveti\v c,  L. Everett, and J. Wang, hep-ph/9808321.}
\bibitem{majoranamass}{L. J. Hall and M. Suzuki, \NPB{231}{84}{419};
T. Banks et al., \PRD{52}{95}{5319};
R. Hempfling, \NPB{478}{96}{3};
B. de Carlos and P. L. White, \PRD{54}{96}{3427};
A. Yu Smirnov and F. Vissani, \NPB{460}{96}{37};
R, M. Borzumati et al., \PLB{384}{96}{123};
H. P. Nilles and N. Polonsky, \NPB{499}{97}{33};
E. Nardi, \PRD{55}{97}{5772}. }

\bibitem{mbmtau}{P. Langacker and N. Polonsky, {\it Phys. Rev.} {\bf D49}
(1994) 1454;
M. Carena {\it et al.}, {\it Nucl. Phys.} {\bf B426} (1994) 269;
V. Barger, M.S. Berger and P. Ohmann, {\it Phys. Rev.} {\bf D47} (1993)
1093.}
\bibitem{Mohap}{R. Mohapatra, [hep-ph/9604414].}
\bibitem{lykken}{G. Aldazabal, 
A. Font,
L. E. Ib\'a\~nez, and G. Violero, [hep-th/9804026]; Z. Kakushadze,
[hep-th/9804110]; J. Lykken, E.  Poppitz
and S. P. Trivedi, [hep-th/9806080].}  }
\end{references}

\newpage

\begin{center}
\begin{tabular}{|c|c|c|ccccccc|c|c|}
\hline\hline
$(SU(3)_C,SU(2)_L,$&
    &CHL
&$Q_1$&$Q_2$&$Q_3$&$Q_4$&$Q_5$&$Q_6$&$Q_A$
                  &$6Q_{Y}$&$100Q_{Y'}$\\
$SU(4)_2,SU(2)_2)$ &&      &&&&&&&&&\\
\hline\hline
(3,2,1,1):&$Q_a$& $Q_1$ & $-$2&0&8&$-$2&$-$8&16&$-$16&  1&68\\
&$Q_b$ & $Q_2$&2&0&8&$-$2&$-$8&16&$-$16&  1&68\\
&$Q_c$ & $Q_3$&0&$-$2&0&$-$2&20&4&$-$12&  1&$-$71\\
\hline
($\bar{3}$,1,1,1):&$u^c_a$& $u^c_1$&2&0&0&6&0&$-$16&$-$16&  $-$4&6\\
&$u^c_b$ &$u^c_2$&$-$2&0&0&6&0&$-$16&$-$16&  $-$4&6\\
&$u^c_c$ &$u^c_3$&0&6&$-$8&2&$-$4&$-$28&$-$12&  $-$4&$-$133\\
&$d^c_a$ &$d^c_1$&0&$-$6&$-$8&2&$-$4&$-$28&$-$12&  2&$-$3\\
&$d^c_b$ &$d^c_2$&0&0&0&0&$-$12&20&4&  2&136\\
&$d^c_c$ &$d^c_3$&0&2&$-$8&$-$2&0&8&8&  2&$-$3\\
&$d^c_d$ &$\bar{t}$&0&2&$-$8&$-$2&0&8&8&  2&$-$3\\
\hline
(1,2,1,1):&$\bar{h}_a$ &$\bar{h}_1$&0&0&$-$8&$-$4&$-$4&$-$12&4&  3&$-$74\\
&$\bar{h}_b$ &$\bar{h}_2$&0&$-$2&$-$16&2&0&0&0&  3&65\\
&$\bar{h}_c$ &$\bar{h}_3$&0&$-$4&8&0&$-$16&24&24&  3&204\\
&$\bar{h}_d$ &$\bar{h}_4$&0&$-$2&0&$-$2&8&24&$-$8&  3&65\\
&$h_a$ &$h_1$&0&0&8&4&4&12&$-$4&  $-$3&74\\
&$h_b$ &$h_2$&0&2&0&2&16&0&0&  $-$3&$-$65\\
&$h_c$ &$h_3$&0&2&0&2&$-$8&$-$24&8&  $-$3&$-$65\\
&$h_d$ &$h_4$&0&2&16&$-$2&0&0&0&  $-$3&$-$65\\
&$h_e$ &$L_1$&0&4&$-$8&0&16&$-$24&$-$24&  $-$3&$-$204\\
&$h_f$ &$L_2$&0&2&16&$-$2&$-$12&$-$12&$-$28&  $-$3&$-$65\\
&$h_g$ &$L_3$&0&2&0&2&16&0&0&  $-$3&$-$65\\
\hline
(3,1,1,1):&${\cal D}_a$&
$t$&0&0&0&0&12&$-$20&$-$4&  $-$2&$-$136\\
\hline
\hline
\end{tabular}
\end{center}
\noindent Table I(a): List of non-Abelian non-singlet observable sector
fields in
the model with their charges under the $U(1)$ gauge groups, hypercharge as defined
in eq.~(\ref{hyp}), and $U(1)'$ as defined in eq.~(\ref{u1p}).
The second column introduces the notation used throughout this paper, and
the third
column the translation to the notation used in~\cite{chl} (CHL).

\newpage
\begin{center}
\begin{tabular}{|c|c|ccccccc|c|c|}
\hline\hline
$(SU(3)_C,SU(2)_L,$&
    
&$Q_1$&$Q_2$&$Q_3$&$Q_4$&$Q_5$&$Q_6$&$Q_A$
                  &$6Q_{Y}$&$100Q_{Y'}$\\
$SU(4)_2,SU(2)_2)$ &      &&&&&&&&&\\
\hline\hline
(1,2,1,2):& $D_{1-4}$&0&0&$-$8&2&$-$4&$-$12&4&  0&0\\
\hline
(1,1,4,1):& $F_{1,2}$&0&0&12&0&12&0&$-$16&  $-$3&$-$65\\
\hline
(1,1,$\bar{4}$,1):& $\bar{F}_{1,2}$&0&0&4&$-$4&$-$4&24&8&  3&65\\
 &$\bar{F}_{3-6}$&0&0&$-$12&0&0&12&$-$20&  3&65\\
 &$\bar{F}_{7,8}$&2&0&4&2&8&$-$12&$-$12&  $-$3&$-$65\\
 &$\bar{F}_{9,10}$&$-$2&0&4&2&8&$-$12&$-$12&  $-$3&$-$65\\
\hline
(1,1,1,2):& $H_{1,2}$&0&$-$2&8&$-$4&$-$12&12&$-$4&  3&65\\
 &$H_{3,4}$&0&$-$4&0&2&$-$8&24&$-$8&  3&204\\
 &$H_{5,7}$&0&2&$-$8&4&12&$-$12&4&  $-$3&$-$65\\
 &$H_{6,8}$&0&0&16&2&$-$16&0&0&  $-$3&74\\
\hline
(1,1,4,2):& $E_{1,2}$&0&0&$-$4&$-$2&4&$-$24&$-$8& 0&$-$139\\ 
 &$E_{3}$&0&$-$2&4&0&4&0&16&0&0\\
 &$E_{4,5}$&0&$-$2&4&0&$-$8&$-$12&$-$12&0&0\\
\hline
(1,1,$\bar{4}$,2):& $\bar{E}_1$&0&2&$-$4&0&$-$4&0&$-$16&0&0\\
\hline
(1,1,6,1):& $S_1$&2&0&$-$8&2&$-$4&$-$12&4&0&0\\
 &$S_2$&0&$-$4&8&0&$-$4&$-$12&4&0&0\\
 &$S_3$&$-$2&0&$-$8&2&$-$4&$-$12&4&0&0\\
 &$S_4$&2&0&8&$-$2&4&12&$-$4&0&0\\
 &$S_5$&0&4&$-$8&0&4&12&$-$4&0&0\\
 &$S_{6,7}$&0&2&0&2&$-$8&24&$-$8&0&139\\
 &$S_8$&$-$2&0&8&$-$2&4&12&$-$4&0&0\\
\hline
(1,1,1,3): &$T_1$&2&4&0&$-$2&8&24&$-$8&0&0\\
 &$T_2$&0&$-$2&8&2&12&36&$-$12&0&139\\
 &$T_3$&$-$2&4&0&$-$2&8&24&$-$8&0&0\\
\hline
\hline
\end{tabular}
\end{center}
\noindent Table I(b): List of non-Abelian 
non-singlet hidden sector fields in the
model with their charges under the $U(1)$ gauge groups,
 hypercharge as defined in
eq.~(\ref{hyp}), and $U(1)'$ as defined in eq.~(\ref{u1p}). (We largely follow
the notation of~\cite{chl} (CHL).)

\newpage

\begin{center}
\begin{tabular}{|c|c|c|ccccccc|c|c|}
\hline\hline
&
    &
&$Q_1$&$Q_2$&$Q_3$&$Q_4$&$Q_5$&$Q_6$&$Q_A$ 
                  &$6Q_{Y}$&$100Q_{Y'}$\\
&&      &&&&&&&&&\\
\hline\hline
$e^c_{a,c}$ &$e^c_{1,3}$&$S_{25,25'}$&0&$-$2&$-$8&$-$6&4&12&$-$4&6&$-$9\\
$e^c_b$  &$e^c_2$&$S_{10}$&4&$-$2&$-$24&$-$2&$-$4&$-$12&4&6&$-$9\\
$e^c_{d,g}$ &$s_{1,4}$&$S_{21,21'}$&0&$-$4&0&$-$4&$-$8&24&$-$8&6&130\\
$e^c_e$ &$s_2$&$S_{13}$&2&0&$-$16&$-$2&$-$32&0&0&6&130\\
$e^c_f$ &$s_3$&$\bar{S}_{2}$&0&$-$4&$-$16&0&8&24&$-$8&6&130\\
$e^c_h$ &$s_5$&$S_{17}$&$-$2&0&$-$16&$-$2&$-$32&0&0&6&130\\
$e^c_i$ &$s_6$&$S_{9}$&$-$4&$-$2&$-$24&$-$2&$-$4&$-$12&4&6&$-$9\\
$e_{a,b}$ &$\bar{s}_{1,2}$&$S_{16,16'}$&0&4&16&0&16&0&0&$-$6&$-$130\\
$e_c$ &$\bar{s}_3$&$S_{2}$&0&4&16&0&$-$8&$-$24&8&$-$6&$-$130\\
$e_{d,e}$ &$\bar{s}_{4,5}$&$S_{24,24'}$&0&2&24&2&4&12&$-$4&$-$6&9\\
$e_f$ &$\bar{s}_6$&$S_{7}$&0&$-$2&8&2&12&$-$60&20&$-$6&$-$269\\
$\varphi_{1}$ &$\varphi_1$&$S_{4}$&4&0&0&0&0&0&0&0&0\\
$\varphi_{2,3}$ &$\varphi_{2,3}$&$S_{5,5'}$&2&4&0&$-$2&$-$16&0&0&0&0\\
$\varphi_{4,5}$ &$\varphi_{4,5}$&$S_{14,14'}$&2&$-$4&16&$-$2&0&0&0&0&0\\
$\varphi_{6,7}$ &$\varphi_{6,7}$&$S_{6,6'}$&2&$-$4&0&2&16&0&0&0&0\\
$\varphi_{8,9}$ &$\varphi_{8,9}$&$S_{15,15'}$&0&0&$-$16&4&16&0&0&0&0\\
$\varphi_{10,11}$&$\varphi_{10,11}$&$\bar{S}_{6,6'}$ 
&$-$2&4&0&$-$2&$-$16&0&0&0&0\\
$\varphi_{12,13}$&$\varphi_{12,13}$&$S_{18,18'}$ 
&$-$2&$-$4&16&$-$2&0&0&0&0&0\\
$\varphi_{14,15}$&$\varphi_{14,15}$&$\bar{S}_{5,5'}$ 
&$-$2&$-$4&0&2&16&0&0&0&0\\
$\varphi_{16}$ &$\varphi_{16}$&$\bar{S}_{4}$&$-$4&0&0&0&0&0&0&0&0\\
$\varphi_{17}$ &$\chi_1$&$S_{22}$&2&4&0&$-$2&8&24&$-$8&0&0\\
$\varphi_{18,19}$
&$\chi_{2,3}$&$S_{11,11'}$&0&2&8&$-$6&$-$4&$-$12&4&0&$-$139\\
$\varphi_{20,21}$&$\chi_{4,5}$&$S_{19,19'}$ 
&0&2&$-$8&$-$2&0&$-$24&$-$24&0&$-$139\\
$\varphi_{22}$ &$\chi_6$&$S_{12}$&0&2&$-$24&2&28&$-$12&4&0&$-$139\\
$\varphi_{23}$ &$\chi_7$&$\bar{S}_{3}$&0&0&16&$-$4&8&24&$-$8&0&0\\
$\varphi_{24}$ &$\chi_8$&$\bar{S}_{1}$&0&0&0&0&$-$12&$-$12&$-$28&0&0\\
$\varphi_{25}$ &$\chi_9$&$S_{20}$&0&$-$2&8&2&12&36&$-$12&0&139\\
$\varphi_{26}$ &$\chi_{10}$&$S_{23}$&$-$2&4&0&$-$2&8&24&$-$8&0&0\\
$\varphi_{27}$ &$\chi_{11}$&$S_{1}$&0&0&0&0&12&12&28&0&0\\
$\varphi_{28,29}$&$\chi_{12,13}$&$S_{8,8'}$&0&0&0&0&$-$24&$-$24&8&0&0\\
$\varphi_{30}$ &$\chi_{14}$&$S_{3}$&0&0&$-$16&4&$-$8&$-$24&8&0&0\\
\hline
\hline
\end{tabular}
\end{center}
\noindent Table I(c): List of non-Abelian singlet fields in
the model with their charges under the $U(1)$ gauge groups, hypercharge as defined
in eq.~(\ref{hyp}), and $U(1)'$ as defined in eq.~(\ref{u1p}).
The first column gives the notation used throughout this paper, the second
column the translation to the notation used in \cite{chl} (CHL), and the third column
the translation to the notation used in \cite{cceel2}.

\newpage

\vskip 1.truecm 

\begin{center}
\begin{tabular}{|l|c|c|}
\hline\hline
FLAT DIRECTION    &   Dim.  & $\#$ $U(1)$'s  \\
\hline\hline
$P_{1} =\langle \varphi_{28}, \varphi_{27}^2\rangle$    & 0 & 1 \\
$P_{1}'=\langle \varphi_{29}, \varphi_{27}^2\rangle$    & 0 & 1 \\
$P_{2}'=$$\langle 
\varphi_{5},\varphi_{10},\varphi_{30},\varphi_{27}^2\rangle$ & 0 & 3 \\
$P_{2}''=$$\langle 
\varphi_{5},\varphi_{11},\varphi_{30},\varphi_{27}^2\rangle$ & 0 & 3 \\
$P_{3}'=\langle
\varphi_{13},\varphi_{2},\varphi_{30},\varphi_{27}^2\rangle$ & 0 & 3 \\
$P_{3}''=\langle
\varphi_{13},\varphi_{3},\varphi_{30},\varphi_{27}^2\rangle$ & 0 & 3 \\
$P_2P_3|_F=\langle \varphi_{12},\varphi_{10},\varphi_{4},\varphi_{2}, 
\varphi_{30}^2,\varphi_{27}^4 \rangle |_F $& 0 & 4 \\
$P_2'''P_3'''|_F= \langle \varphi_{12},\varphi_{11},\varphi_{4},\varphi_{3}, 
\varphi_{30}^2,\varphi_{27}^4 \rangle |_F $& 0 & 4 \\
$P_{1}P_1'=\langle \varphi_{28},\varphi_{29}, \varphi_{27}^4\rangle$    & 1
& 1 \\
$P_{2}'P_2''=\langle \varphi_{5}^2,\varphi_{10}, 
\varphi_{11},\varphi_{30}^2,\varphi_{27}^4 \rangle$ & 1 & 3 \\
$P_{3}'P_3''=\langle \varphi_{2}^2,\varphi_{12},
\varphi_{13},\varphi_{30}^2,\varphi_{27}^4 \rangle$ & 1 & 3\\
$P_1P_2'=\langle \varphi_{28},\varphi_{5},
\varphi_{10},\varphi_{30},\varphi_{27}^4 \rangle$& 1 & 3\\
$P_1P_2''=\langle \varphi_{28},\varphi_{5},
\varphi_{11},\varphi_{30},\varphi_{27}^4 \rangle$& 1 & 3\\
$P_1'P_2'=\langle \varphi_{29},\varphi_{5},
\varphi_{10},\varphi_{30},\varphi_{27}^4 \rangle$& 1 & 3\\
$P_1'P_2''=\langle \varphi_{29},\varphi_{5},
\varphi_{11},\varphi_{30},\varphi_{27}^4 \rangle$& 1 & 3\\
$P_1P_3'=\langle \varphi_{28},\varphi_{2},
\varphi_{13},\varphi_{30},\varphi_{27}^4 \rangle$& 1 & 3\\
$P_1P_3''=\langle \varphi_{28},\varphi_{3},
\varphi_{13},\varphi_{30},\varphi_{27}^4 \rangle$& 1 & 3\\
$P_1'P_3'=\langle \varphi_{29},\varphi_{2},
\varphi_{13},\varphi_{30},\varphi_{27}^4 \rangle$& 1 & 3\\
$P_1'P_3''=\langle \varphi_{29},\varphi_{3},
\varphi_{13},\varphi_{30},\varphi_{27}^4 \rangle$& 1 & 3\\
$P_2'P_3'=\langle \varphi_{2},\varphi_{5},
\varphi_{10},\varphi_{13},\varphi_{30}^2,\varphi_{27}^4 \rangle$ & 1 & 4\\
$P_2'P_3''=\langle \varphi_{3},\varphi_{5},
\varphi_{10},\varphi_{13},\varphi_{30}^2,\varphi_{27}^4 \rangle$ & 1 & 4\\
$P_2''P_3'=\langle \varphi_{2},\varphi_{5},
\varphi_{11},\varphi_{13},\varphi_{30}^2,\varphi_{27}^4 \rangle$ & 1 & 4\\
$P_2''P_3''=\langle \varphi_{3},\varphi_{5},
\varphi_{11},\varphi_{13},\varphi_{30}^2,\varphi_{27}^4 \rangle$ & 1 & 4
\\
$P_1P_2P_3|_F=\langle \varphi_{12},\varphi_{10},\varphi_{4},\varphi_{2},
\varphi_{28},\varphi_{30}^2,\varphi_{27}^6 \rangle |_F $ & 1 & 4\\
$P_1'P_2P_3|_F=\langle \varphi_{12},\varphi_{10},\varphi_{4},\varphi_{2},
\varphi_{29},\varphi_{30}^2,\varphi_{27}^6 \rangle |_F $ & 1 & 4\\
$P_1P_2'''P_3'''|_F=\langle
\varphi_{12},\varphi_{11},\varphi_{4},\varphi_{3},
\varphi_{28},\varphi_{30}^2,\varphi_{27}^6 \rangle |_F $ & 1 & 4\\
$P_1'P_2'''P_3'''|_F=\langle
\varphi_{12},\varphi_{11},\varphi_{4},\varphi_{3},
\varphi_{29},\varphi_{30}^2,\varphi_{27}^6 \rangle |_F$& 1 &4\\
$P_1P_2'P_3'=\langle
\varphi_{13},\varphi_{10},\varphi_{5},\varphi_{2},
\varphi_{28},\varphi_{30}^2,\varphi_{27}^6 \rangle $& 2 &4\\
$P_1'P_2'P_3'=\langle
\varphi_{13},\varphi_{10},\varphi_{5},\varphi_{2},
\varphi_{29},\varphi_{30}^2,\varphi_{27}^6 \rangle $& 2 &4\\
$P_1P_2'P_3''=\langle
\varphi_{13},\varphi_{10},\varphi_{5},\varphi_{3},
\varphi_{28},\varphi_{30}^2,\varphi_{27}^6 \rangle $& 2 &4\\
$P_1'P_2'P_3''=\langle
\varphi_{13},\varphi_{10},\varphi_{5},\varphi_{3},
\varphi_{29},\varphi_{30}^2,\varphi_{27}^6 \rangle $& 2 &4\\
$P_1P_2''P_3'=\langle
\varphi_{13},\varphi_{11},\varphi_{5},\varphi_{2},
\varphi_{28},\varphi_{30}^2,\varphi_{27}^6 \rangle $& 2 &4\\
$P_1'P_2''P_3'=\langle
\varphi_{13},\varphi_{11},\varphi_{5},\varphi_{2},
\varphi_{29},\varphi_{30}^2,\varphi_{27}^6 \rangle $& 2 &4\\
$P_1P_2''P_3''=\langle
\varphi_{13},\varphi_{11},\varphi_{5},\varphi_{3},
\varphi_{28},\varphi_{30}^2,\varphi_{27}^6 \rangle $& 2 &4\\
$P_1'P_2''P_3''=\langle
\varphi_{13},\varphi_{11},\varphi_{5},\varphi_{3},
\varphi_{29},\varphi_{30}^2,\varphi_{27}^6 \rangle $& 2 &4\\
\hline
\hline
\end{tabular}
\end{center}
\noindent Table II:  The explicit list of type-B $D$- flat directions that are
$F$- flat to all orders for the CHL5 model. (This table expands the compact
presentation of these flat directions  given in ~\cite{cceel2}.)
 The dimension of the
direction, after
cancellation of the Fayet-Iliopoulos  term, is indicated in the second
column. The third column gives the number (out of six) of non-anomalous $U(1)$'s
broken along the flat direction. 

\newpage
\vskip 1.truecm

\begin{center}
\begin{tabular}{|c|c|}
\hline\hline
Massive Fields    &   Mass \\
\hline\hline
${\bar h}_b$,$h_f'=\frac{1}{\sqrt{N_1}}[\sqrt{2}gx
h_f+\frac{(\alpha^{(1)}_4 |\psi_2|^2+\alpha^{(2)}_4
(|\psi_1|^2-|\psi_2|^2)
}{M_{Pl}}h_b]$&$\sqrt{N_1}$\\
${h}_g$,$\bar{h}_d$&$g\sqrt{x^2-|\psi_1|^2}$ \\
$e_d^c$, $e_g^c$, $e_a$, $e_b$ & ${g\over \sqrt{2}}|\psi_1|$\\
$\varphi_9,\varphi_4'=\frac{1}{|\psi_1|}(|\psi_2|\varphi_4+\sqrt{|\psi_1|^2- 
|\psi_2|^2}\varphi_{12})$&${{g\over \sqrt{2}}}|\psi_1|$\\
$\varphi_1$, $\varphi_{15}$&${{g\over \sqrt{2}}}|\psi_2|$\\
$\varphi_{7}$,
$\varphi_{16}$&${g\over \sqrt{2}}\sqrt{|\psi_1|^2-|\psi_2|^2}$\\
$\varphi_{6}, \varphi_{8}, \varphi_{14}, \varphi_{17}, \ 
\varphi_{23},  \varphi_{26}$&${g\over \sqrt{2}}{\sqrt{x^2-|\psi_1|^2}}$\\
$\varphi_{21}$,
$\varphi_{25}$&$\sqrt{N_2}$\\
$F_1,F_2,\bar{F}_1,\bar{F}_2$
&${g\over \sqrt{2}}|\psi_1|$\\
$S_5$,$S_3'=\frac{1}{|\psi_1|}(|\psi_2|S_3+\sqrt{|\psi_1|^2-|\psi_2|^2}S_1)$
&${g\over \sqrt{2}}|\psi_1|$\\\hline  
\hline
\end{tabular}  
\end{center}
\noindent Table IIIa: List of massive states for the $P_1'P_2'P_3'$
flat direction as can be read off from  the effective bilinear
superpotential
terms (\ref{p123massw}). Here  $N_1\equiv 
2g^2x^2+[\alpha^{(1)}_4|\psi_2|^2+\alpha^{(2)}_4
(|\psi_1|^2-|\psi_2|^2)]^2/M_{Pl}^2$ and
$N_2 \equiv 2(\alpha_4^{(3)}/M_{pl})^2x^2(x^2-|\psi_1|^2)$ and  the
definition of VEV
parameters $x, \  \psi_1$ and $\psi_2$ is given in eqs. (\ref{vevs}).
The  fields  with nonzero VEV's in the flat directions
($\varphi_2,\ \varphi_5, \ \varphi_{10}, \ \varphi_{13}, \ \varphi_{27},
\ \varphi_{29}, \ \varphi_{30}$) are discussed in
the text (at the  end of Section IIB) and are not given in this Table.
[Five
complex fields contribute to the Higgs mechanism,
completing five massive vector supermultiplets, associated with the
spontaneous symmetry breaking of five $U(1)$ factors (including anomalous
$U(1)$) and two complex fields remain massless (and act as moduli
associated
with the two free parameters of VEV's).]

\newpage
\begin{center}
\begin{tabular}{|c|}
\hline\hline
Massless Fields\\  
\hline\hline
 $Q_{a}$,$Q_{b}$,$Q_{c}$\\
$u^c_{a}$,$u^c_{b}$,$u^c_{c}$\\
$d^c_{a}$,$d^c_{b}$,$d^c_{c}$,$d^c_{d}$,${\cal D}_a$\\
$\bar{h}_a$,$\bar{h}_c$\\
$h_a$,$h_c$,$h_d$,$h_e$\\
$h_b'=\frac{1}{\sqrt{N_1}}[-\frac{\alpha^{(1)}_4
|\psi_2|^2+\alpha^{(2)}_4 (|\psi_1|^2-|\psi_2|^2)}{M_{Pl}}
h_f+\sqrt{2}gx h_b$]\\
$e^c_a,e^c_b,e^c_c,e^c_e,e^c_f,e^c_h,e^c_i$\\
$e_c,e_d,e_e,e_f$\\
$\varphi_{3},\varphi_{11},$
$\varphi_{18},\varphi_{19},\varphi_{20},\varphi_{22},\varphi_{24},\varphi_{28}
$\\
$\varphi_{12}'=\frac{1}{|\psi_1|}(-\sqrt{|\psi_1|^2-|\psi_2|^2}\varphi_{12} 
+|\psi_2|\varphi_4)$\\
$D_1,D_2,D_3,D_4$\\
$\bar{F}_3,\bar{F}_4,\bar{F}_5,\bar{F}_6,\bar{F}_7,
\bar{F}_8,\bar{F}_9,\bar{F}_{10}$\\
$H_{1},H_{2},H_{3},H_{4},H_{5},H_{6},H_{7},H_{8}$\\
$\bar{E}_{1},E_1,E_2,E_3,E_4,E_{5}$\\
$S_{2},S_{4},S_{6},S_{7},S_8$\\
$S_1'=\frac{1}{|\psi_1|}(-\sqrt{|\psi_1|^2-|\psi_2|^2}S_3+|\psi_2|S_1)$\\$T_{1},T_{2},T_{3}$\\
\hline
\hline
\end{tabular}
\end{center}
\noindent Table IIIb: List of massless states (excluding the two moduli) 
for the $P_1'P_2'P_3'$
flat direction. These are fields without an  effective
bilinear term in
the superpotential  (\ref{p123massw}).
\newpage

\newpage
\begin{center}
\begin{tabular}{|c|c|}
\hline\hline
Massive Fields    &   Mass \\
\hline\hline
$\varphi_9$ & $gx$\\
$\bar{h}_b$,$h_f$ & $\sqrt{2}gx$ \\
$e_d^c$, $e_g^c$, $e_a$, $e_b$ & ${{g\over \sqrt{2}}}x$\\
$\varphi_{8}$,
$\varphi_{3'}=(\varphi_{3}+\varphi_{11})/\sqrt{2}$ &
${g\over \sqrt{2}}x$ \\ 
$\varphi_{15}$,
$\varphi_{1'}=(-\frac{gx}{{2}}\varphi_{1}+
\frac{\alpha^{(1)}_{5}}{M_{Pl}^{2}}\frac{x^{3}}{2}\varphi_{26})/\sqrt{N_1}$ 
& $\sqrt{N_1}$ \\
$\varphi_{7}$,
$\varphi_{16'}=(\frac{gx}{{2}}\varphi_{16}-
\frac{\alpha^{(2)}_{5}}{M_{Pl}^{2}}\frac{x^{3}}{2}\varphi_{17})/\sqrt{N_2}$
& $\sqrt{N_2}$ \\
$F_1$, $F_2$, $\bar{F}_1$, $\bar{F}_2$ & ${g\over\sqrt{2}}x$\\
\hline  
\hline
\end{tabular}  
\end{center}
\noindent Table IVa: List of massive states for the $P_2P_3|_F$
flat direction, with
$N_{1/2}\equiv g^2x^{2}/4+(\alpha_5^{(1/2)}x^3/2M_{Pl}^{2})^2$. 
Except for $\varphi_9$ the effective bilinear terms can be read off eq. 
(\ref{p23Fmassw}). The fields with
nonzero VEV's in the flat directions ($\varphi_2,\ \varphi_4, \
\varphi_{10}, \ \varphi_{12}, \ \varphi_{27},\ \varphi_{30}$) are
discussed in
the text (at end of Section IIB) and are not given in this Table. [Five
complex fields contribute to the Higgs mechanism,
completing  five massive vector supermultiplets, associated with the
spontaneous symmetry breaking of five $U(1)$ factors (including anomalous
$U(1)$) and one complex field gets its mass due to  the superpotential 
terms which impose  $F$- flatness constraints on  VEV's, i.e.,
$(gx/{2})\varphi_9(\delta\varphi_{2}+\delta\varphi_{12}-
\delta\varphi_{4}+\delta\varphi_{10})$.
Note that $\varphi_9$ acquires
 mass due to  the same  coupling in
the superpotential.]
\newpage
\begin{center}
\begin{tabular}{|c|}
\hline\hline
Massless Fields\\  
\hline\hline
 $Q_{a}$,$Q_{b}$,$Q_{c}$\\
$u^c_{a}$,$u^c_{b}$,$u^c_{c}$\\
$d^c_{a}$,$d^c_{b}$,$d^c_{c}$,$d^c_{d}$,${\cal D}_a$\\
$\bar{h}_a$,$\bar{h}_c$, $\bar{h}_d$ \\
$h_a$,$h_b$,$h_c$,$h_d$,$h_e$, $h_g$ \\
$e_a^c$, $e_b^c$, $e_c^c$, $e_e^c$, $e_f^c$, $e_h^c$, $e_i^c$ \\
$e_c$, $e_d$, $e_e$, $e_f$ \\
$\varphi_{5}, \varphi_{6}, \varphi_{13}, \varphi_{14},
\varphi_{18}, \varphi_{19}, \varphi_{20}, \varphi_{21}, \varphi_{22}, 
\varphi_{23}, \varphi_{24}, \varphi_{25},\varphi_{28},
\varphi_{29}$ \\
$\varphi_{11'}=(\varphi_{3}-\varphi_{11})/\sqrt{2}$, \\
$\varphi_{17'}=(\frac{gx}{{2}}\varphi_{17}+\frac{\alpha^{(2)}_{5}}{M_{Pl}^{2}}
\frac{x^{3}}{2}\varphi_{16})/\sqrt{N_2}$\\
$\varphi_{26'}=(-\frac{gx}{{2}}\varphi_{26}-\frac{\alpha^{(1)}_{5}}
{M_{Pl}^{2}}\frac{x^{3}}{2}\varphi_{1})/\sqrt{N_1}$\\
$D_1,D_2,D_3,D_4$\\
$\bar{F}_3,\bar{F}_4,\bar{F}_5,\bar{F}_6,\bar{F}_7,
\bar{F}_8,\bar{F}_9,\bar{F}_{10}$\\
$H_{1},H_{2},H_{3},H_{4},H_{5},H_{6},H_{7},H_{8}$\\
$\bar{E}_{1},E_1,E_2,E_3,E_4,E_{5}$\\
$S_1,S_{2},S_3,S_{4},S_5,S_{6},S_{7},S_8$\\
$T_{1},T_{2},T_{3}$\\
\hline\hline
\end{tabular}
\end{center}
\noindent Table IVb: List of massless states (without effective bilinear 
terms in (\ref{p23Fmassw})) for the $P_2P_3|_F$
flat direction. 

\vskip 1.truecm

\begin{center}
\begin{tabular}{|c|c|c|c|c|c|c|c|}
\hline\hline

 {\rm Effective 
$\beta$} & $\beta_1$ & $\beta_2$ & $\beta_3$ &
$\beta_{1^{'}}$ & $\beta_{11^{'}}$ & $\beta_{2hid}$ & $\beta_{4hid}$ \\
\hline \hline
$P_1^{'}P_2^{'}P_3^{'}$ Flat Direction & $10.0$ & $6.0$ & $-2.0$ & $10.2$ & $4.8$ & $10.0$ & $2.0$ \\
$P_2P_3|_F$ Flat Direction & $10.3$ & $7.0$ & $-2.0$ & $10.6$ & $5.0$ & $10.0$ & $3.0$ \\
\hline \hline
\end{tabular}
\end{center}
\noindent Table V: Effective beta-functions are quoted for the
two representative
flat directions.  The effective  beta-function is defined as  $\beta_{i}\equiv
\beta^0_i/k_i$, where  $\beta^0_i$  and $k_i$ are the   beta-function
and the
Ka\v c-Moody level for  a particular gauge group factor, respectively.
 The subscripts $1,\ 2, \ 3, \ 1', \  {2hid},
\ {4hid}$  refer to $U(1)_Y, \ SU(2)_L, \ SU(3)_C, \ U(1)',\ SU(2)_2, \
SU(4)_2$ gauge group factors and $11'$ refers to the $U(1)_Y$ and $U(1)'$
kinetic  mixing. The Ka\v c-Moody levels are $k_1=11/3$, $k_2=k_3=1$,
$k_{1'}\simeq
16.67$,  and  $k_{2hid}=k_{4hid}=2$.





\end{document}